\documentclass[secnumarabic,amssymb,nobibnotes,aps,prl,preprint,letterpaper,preprintnumbers,amsmath,superscriptaddress,floatfix]{revtex4-1}

\usepackage{wrapfig}
\usepackage{amsmath}
\usepackage{enumitem}
\usepackage{multirow}
\usepackage[letterpaper,margin={1in,1in}]{geometry}
\usepackage{graphicx}
\usepackage{epstopdf}
\usepackage{lineno}
\usepackage{appendix}
\usepackage{longtable}
\usepackage{array}
\usepackage{siunitx}
\usepackage[dvipsnames]{xcolor}
\setcitestyle{super}
\usepackage{multirow}
\DeclareSIUnit\mrad{\milli\rad}
\usepackage{color}
\usepackage{xcolor}

\begin{document}


\title{Origin and stability of the charge density wave in ScV$_6$Sn$_6$}

\author{Yanhong Gu}
\address{Department of Chemistry, University of Tennessee, Knoxville, Tennessee 37996, USA}

\author{Ethan Ritz}
\address{Department of Chemical Engineering and Materials Science,
University of Minnesota, Minneapolis, Minnesota 55455, USA}

\author{William R. Meier}
\address{Materials Science and Engineering Department, University of Tennessee Knoxville, Knoxville, Tennessee 37996, USA}

\author{Avery Blockmon}
\address{Department of Chemistry, University of Tennessee, Knoxville, Tennessee 37996, USA}

\author{Kevin Smith}
\address{Department of Chemistry, University of Tennessee, Knoxville, Tennessee 37996, USA}

 \author{Richa Pokharel Madhogaria}
\address{Materials Science and Engineering Department, University of Tennessee Knoxville, Knoxville, Tennessee 37996, USA}
 
 \author{Shirin Mozaffari} 
\address{Materials Science and Engineering Department, University of Tennessee Knoxville, Knoxville, Tennessee 37996, USA}

\author{David Mandrus}
\address{Materials Science and Engineering Department, University of Tennessee Knoxville, Knoxville, Tennessee 37996, USA}
\address{Department of Physics and Astronomy, University of Tennessee Knoxville, Knoxville, Tennessee 37996, USA}
\address{Materials Science and Technology Division, Oak Ridge National Laboratory, Oak Ridge, Tennessee 37831, USA}

\author{Turan Birol}
\address{Department of Chemical Engineering and Materials Science,
University of Minnesota, Minneapolis, Minnesota 55455, USA}

\author{Janice L. Musfeldt}
 \email{musfeldt@tennessee.edu}
\address{Department of Chemistry, University of Tennessee, Knoxville, Tennessee 37996, USA}
\address{Department of Physics and Astronomy, University of Tennessee, Knoxville, Tennessee 37996, USA}

\date{\today}

\begin{abstract}

{\bf Kagom\'e metals are widely recognized 
as versatile platforms for exploring 
novel topological properties, unconventional electronic correlations, magnetic frustration, and superconductivity.
In the $R$V$_6$Sn$_6$ family of materials ($R$ = Sc, Y, Lu),  ScV$_6$Sn$_6$  
hosts an unusual charge density wave ground state as well as structural similarities  
with the $A$V$_3$Sb$_5$ system ($A$ = K, Cs, Rb). 
In this work, we combine Raman scattering spectroscopy with first-principles lattice dynamics calculations to reveal the charge density wave state in ScV$_6$Sn$_6$. In the low temperature phase, we find a five-fold splitting of the V-containing totally symmetric mode near 240 cm$^{-1}$ suggesting that the 
 density wave acts to mix modes of $P$6/$mmm$ and $R$$\bar{3}$$m$  symmetry - an effect that we quantify by projecting phonons of the high symmetry state onto those of the lower symmetry structure. 
 We also test the stability of the density wave state under compression and find that both physical and chemical pressure act to quench the effect. 
We discuss these findings in terms of symmetry and the structure-property trends that can be unraveled in this system.}


\end{abstract}

\maketitle


\section*{Introduction}

Kagom\'e materials possess two-dimensional periodic networks of corner-sharing triangles and, as a result, exhibit a high degree of geometrical frustration.
This structural frustration can create 
Dirac cones and flat bands in the electronic band structure, as well as exotic magnetic ground states \cite{kang2020dirac,li2021dirac,liu2020orbital,yin2019negative,balents2010spin,yan2011spin,William2020, paul2020spin}  accompanied by the anomalous Hall effect, \cite{ye2018massive,lachman2020exchange,tanaka2020topological,xu2015intrinsic}  charge fractionalization,\cite{ruegg2011fractionally,feng2017gapped}  chiral magnetism, \cite{ghimire2020competing,Wang2021prb} and strong electron correlations.\cite{teng2023magnetism}  
The discovery of charge density waves (CDWs) in superconducting $A$V$_3$Sb$_5$ ($A$ = K, Rb, Cs) and magnetic  FeGe demonstrates that CDWs can exist in both magnetic and nonmagnetic kagom\'e lattices across a range of electron correlations. \cite{ortiz2019prm,Ortiz20220prl,teng2022nature}  
At the same time, rich phase diagrams can be obtained by tuning the frustration and electron filling in the kagom\'e lattice. Strong entanglements make this platform well-suited to revealing intertwined and competing states.    
%
Recently, a family of bi-layer analogs with chemical formula $R$V$_6$Sn$_6$ ($R$ = Sc, Y, Lu, Tb, Ho, Gd...) has attracted attention [Fig. \ref{fig:1}\textbf{a}].\cite{EPokharel2021prb} 
These kagom\'e metals also host topological Dirac surface states, van Hove singularies, anisotropic magnetism, and other exciting properties.\cite{hu2022tunable,zhang2022prm}
That said, unlike all three $A$V$_3$Sb$_5$ compounds, superconductivity has not been reported in ScV$_6$Sn$_6$ under any temperature or pressure conditions investigated thus far, \cite{zhang2022destabilization,arachchige2022charge} possibly due to the lack of a $\Gamma$-centered Fermi pocket in this material. \cite{ritz2023orbital,zhao2021cascade,Liang2021prx,Feng2021prb} 
Also, remarkably, only ScV$_6$Sn$_6$ exhibits a three dimensional CDW, making it comparable to the $A$V$_3$Sb$_5$'s. Despite recent activity in this highly contemporary research area, there is much more to learn about the CDW in ScV$_6$Sn$_6$ and related materials. 
In particular, the combination of two kagom\'e layers per unit cell, along with the lower symmetry wavevector of the CDW makes resolving structural details and phase transition of ScV$_6$Sn$_6$ a particularly daunting task.

ScV$_6$Sn$_6$ is a paramagnetic metal with a first-order CDW transition at 92 K ($T_{\rm CDW}$).\cite{Tianchen2023prb,arachchige2022charge,cao2023competing, korshunov2023softening} The CDW primarily involves out-of-plane Sc and Sn displacements; thus far, the contribution of V centers has been neglected.\cite{arachchige2022charge, tan2023abundant}  
First-principles calculations point to lattice instabilities (soft modes) in ScV$_6$Sn$_6$\cite{tan2023abundant}  -  possibly due to the small size of the Sc$^{3+}$ radius. Experimentally, a phonon mode leads to short range order in the high temperature phase with wavevector (1/3, 1/3, 1/2). A stable long range CDW order with wavevector (1/3, 1/3, 1/3) sets in below 92 K in the low temperature phase.\cite{cao2023competing, korshunov2023softening}
%
This type of lattice instability is not present in the Y and Lu analogs, and there are no CDWs in these materials.\cite{tan2023abundant} 
Experimental confirmation of these predictions by other probes is highly desirable. 
Traditionally, x-ray techniques and vibrational spectroscopy have been favored for unraveling these issues. Infrared and Raman scattering spectroscopies in particular are well-suited for exploring the microscopic aspects of local lattice distortions as well as phase and amplitude modes of a CDW. Of course, the metallic character of ScV$_6$Sn$_6$ challenges this approach because the odd-symmetry infrared-active phonons are screened by the Drude contribution.\cite{Fan2017} Raman scattering  provides a way forward,\cite{Fan2020} and it has been used to gain significant insight about the CDW in other kagom\'e systems\cite{wu2022charge, liu2022observation} even though it accesses only even-symmetry features at the zone center.
Compared to the AV$_3$Sb$_5$ family, Raman spectroscopy can provide more information about ScV$_6$Sn$_6$ because of the larger number of Raman active lattice modes in its crystal structure.

In this work, we combine temperature- and pressure-dependent Raman scattering spectroscopy of ScV$_6$Sn$_6$ with complementary lattice dynamics calculations to reveal the properties of the charge density wave states in this model bi-layer kagom\'e metal. What distinguishes our work from prior efforts\cite{hu2023phonon} is the quality of our single crystals with different $R$ site substitutions, the ability to employ both temperature and pressure as tuning parameters, and our symmetry-guided strategy of projecting the high temperature phase $P$6/$mmm$ phonons onto those in the low temperature CDW state to uncover their origins. We find that the $A_{1g}$ symmetry mode near 240 cm$^{-1}$ - which involves out-of-plane V center motion - is very sensitive to the development of the CDW. For instance, even though it corresponds to a nondegenerate phonon mode, it seemingly displays five-fold splitting in the low temperature phase consistent with $R\bar{3}m$ symmetry. We discuss this symmetry breaking in terms of mixing of nearby symmetry-appropriate and zone-folded phonons. 
We also demonstrate that compression at room temperature quenches the recently reported short range CDW in ScV$_6$Sn$_6$ but has no effect on the Lu analog. We therefore establish that density wave stability is impacted by both physical and chemical pressure. 
These findings revise our understanding of how and why CDWs are stabilized in ScV$_6$Sn$_6$ and related materials.


\section*{Results and Discussion}

\subsection*{Raman-active phonons of ScV$_6$Sn$_6$ at 300 K}

\begin{figure*}[b!]
\centering
\includegraphics[width=\linewidth]{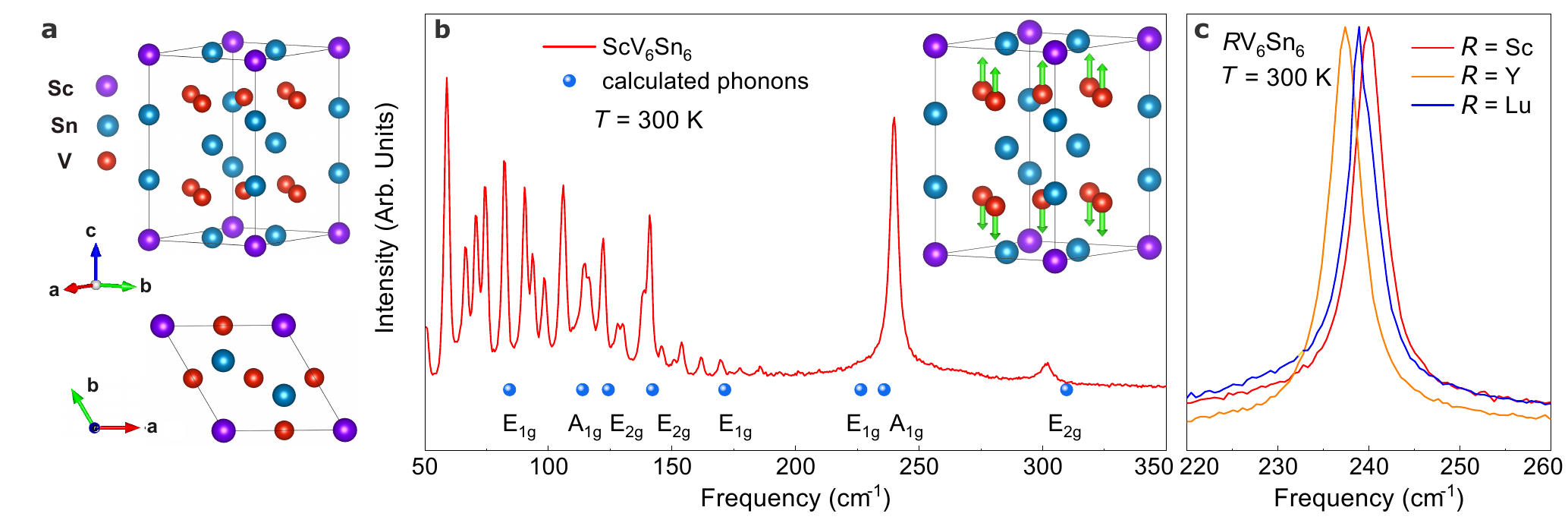}
\caption{ \textbf{Properties of hexagonal ScV$_6$Sn$_6$ at room temperature.} \textbf{a} Crystal structure of ScV$_6$Sn$_6$ (space group $P6/mmm$, \#191). \cite{arachchige2022charge} This three dimensional material hosts two vanadium kagom\'e layers separated by ScSn$_2$ and Sn$_2$ layers along the out of plane direction. \textbf{b} Raman scattering spectrum of ScV$_6$Sn$_6$ compared with predictions from complementary lattice dynamics calculations. The spectrum is collected on the $ab$ plane in the backscattering geometry. 
The inset shows the displacement pattern of the $A_{1g}$ symmetry mode near 240 cm$^{-1}$.
\textbf{c} Close up view of the phonon near 240 cm$^{-1}$ in the $R$V$_6$Sn$_6$ family materials ($R$ = Sc, Y, Lu).}
\label{fig:1}
\end{figure*}

Figure \ref{fig:1}\textbf{b} summarizes the Raman scattering response of ScV$_6$Sn$_6$ at room temperature.  We employ a symmetry analysis, complementary lattice dynamics calculations, polarizer/analyzer combinations, and chemical substitution on the $R$ site to assign the excitations. Overall, the eight calculated phonon frequencies (blue spheres in Fig. \ref{fig:1}\textbf{b}, see Supplementary Table II) are in excellent agreement with the measured spectrum. Challenging assignments are near 240 cm$^{-1}$ where there are two candidate modes and below 200 cm$^{-1}$ where the Raman-active phonons are embedded in a series of extra peaks. The latter turn out to be rotational modes of air in the optical path. They appear in this and other measurements of ScV$_6$Sn$_6$ \cite{ohno2021pure,hu2023phonon} due to the low brightness of the sample. The peak near 240 cm$^{-1}$ is important in our continuing discussion. According to our lattice dynamics calculations, there are two candidates for this structure: $E_{1g}$ and $A_{1g}$ symmetry modes. Surprisingly, the peak can be fit by a single oscillator suggesting that one of these constituents is present with extremely low intensity. Based upon extinction behavior under various polarizer/analyzer arrangements  [Supplementary figure 1 \textbf{c}], we assign the 240 cm$^{-1}$ peak as primarily an $A_{1g}$ symmetry mode. 
The displacement pattern, which involves out-of-plane  V center motion, is shown in the inset of Fig. \ref{fig:1}\textbf{b}. 
Substitution on the $R$ (Sc) site is less useful for assignment purposes because the frequency vs. mass trend is not straightforward [Fig. \ref{fig:1}\textbf{c}]. The atomic radius of the $R$ center and the precise local environment appear to be more important. 
The complete set of mode assignments in ScV$_6$Sn$_6$ is given in Supplementary Table 1.

\subsection*{Symmetry breaking across the CDW transition}

\begin{figure*}[t!]
\centering
\includegraphics[width=\linewidth]{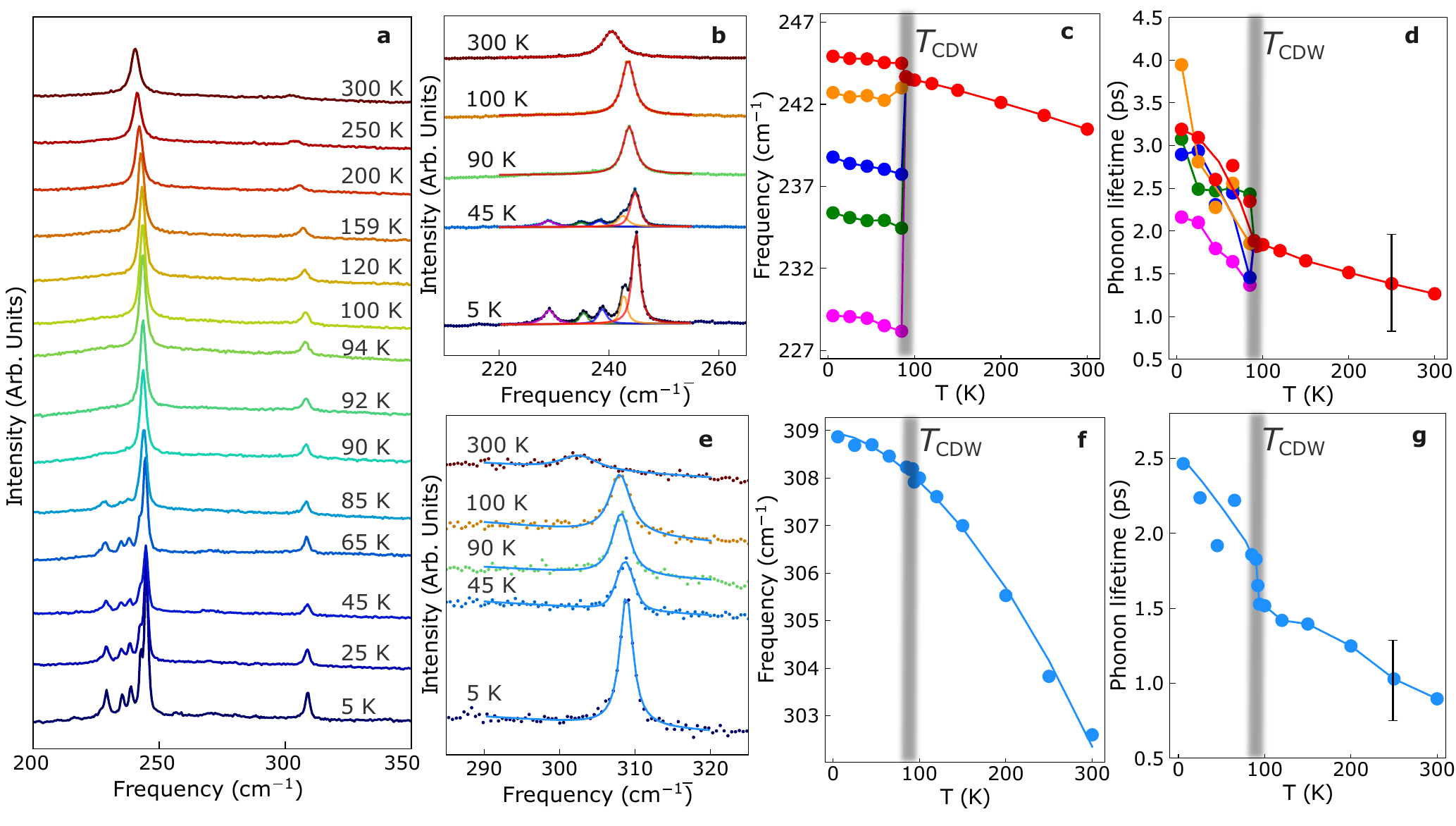}
\caption{\textbf{Symmetry breaking across the CDW transition in ScV$_6$Sn$_6$.} \textbf{a} Summary of the Raman-active modes near 240 and 300 cm$^{-1}$ as a function of temperature. All the spectra are collected on the $ab$ plane in the back-scattering geometry.  \textbf{b} Close-up view of the spectra at selected temperatures along with an oscillator fit.  Several new features emerge in the CDW phase due to symmetry breaking.  A minimum of five Voigt oscillators are needed to model the base temperature data. \textbf{c, d} Frequency vs. temperature and phonon lifetime vs. temperature trends for the features near 240 cm$^{-1}$. These findings were extracted from the results in panel \textbf{b}. In \textbf{c}, error bars are on the order of the symbol size; a characteristic error bar is indicated in panel \textbf{d}.   \textbf{e} Close-up view of the $E_{2g}$ symmetry mode near 300 cm$^{-1}$ as a function of temperature. This mode can be fit with a single Voigt oscillator over the full temperature range.  \textbf{e, f} Frequency vs. temperature and phonon lifetime vs. temperature trends for the $E_{2g}$ symmetry mode near 300 cm$^{-1}$. In \textbf{f}, error bars are on the order of the symbol size; a characteristic error bar is shown in panel \textbf{g}.}
\label{fig:2}
\end{figure*}

Figure \ref{fig:2}\textbf{a} summarizes the Raman scattering response of ScV$_6$Sn$_6$ as a function of temperature in the high frequency region. We focus on the behavior of the $A_{1g}$ symmetry mode near 240 cm$^{-1}$. This structure hardens systematically with decreasing temperature and splits into a cluster of at least five closely-spaced peaks below 90 K. \footnote{The low brightness of this sample and resolution issues make it difficult to tell whether the small shoulders and tiny features on the baseline are real - sufficiently different from nearby peaks as well as the noise level. }
This cluster is the most conspicuous signature of the CDW phase, although it obviously raises questions of exactly how and why a singly-degenerate vibrational mode might split in a low symmetry environment.  
We show the behavior of the $E_{2g}$ phonon near 300 cm$^{-1}$  for comparison. This structure hardens anharmonically with decreasing temperature but does not split across $T_{\rm CDW}$  - in line with the other Raman-active modes in ScV$_6$Sn$_6$ including those of $A_{1g}$ symmetry. 
Complementary measurements of LuV$_6$Sn$_6$ and YV$_6$Sn$_6$ reveal no low temperature splitting of the 240 cm$^{-1}$ mode [Supplemental figure 2], consistent with transport results indicating the absence of CDW transitions in the $R$ = Lu and Y analogs.\cite{zhang2022prm,EPokharel2021prb} This again demonstrates that 
multiplet splitting of the $A_{1g}$ symmetry mode is a signature of the unusual CDW state in ScV$_6$Sn$_6$.

We quantify these results by fitting the Raman scattering response of ScV$_6$Sn$_6$  with a series of Voigt oscillators and a linear baseline.  Close-up views of the modes of interest along with their oscillator fits are shown in Fig. \ref{fig:2}\textbf{b,e}. 
A single oscillator is required to fit the $A_{1g}$ symmetry mode in the normal state whereas a minimum of five oscillators are needed to fit the spectra in the CDW state.  
By contrast, a single oscillator fits the $E_{2g}$ mode near 300 cm$^{-1}$ over the full temperature range.   
The center frequency and full width at half maximum (FWHM) are extracted from these fits.  
The phonon lifetime $\tau$ is a useful quantity related to the Heisenberg uncertainty principle that can be calculated as $\tau = \hbar/ \text{FWHM}$.\cite{sun2013spectroscopic}

Figure \ref{fig:2}\textbf{c,d} displays frequency and phonon lifetime trends as a function of temperature for the $A_{1g}$ symmetry phonon. The sharp peak splitting immediately below $T_{\rm CDW}$ is consistent with a first-order phase transition,\cite{Tianchen2023prb,arachchige2022charge,cao2023competing, korshunov2023softening} although as we shall see below, the splitting is not due to symmetry breaking induced splitting of components of a single mode because a singly-degenerate vibrational mode cannot not split further as part of a group-subgroup transition. 
At the same time, the strong clustering  seems to argue against the appearance of traditional zone-folded phonons unless the phonon bands are rather flat and fold into a similar frequency window. \cite{liu2022observation,Joshi2019prb,Hill2019prb}  
Overall, the phonon lifetime rises from 1.25 ps at room temperature to between 2.2 and 4 ps  in the low temperature phase depending on the branch.

Figure \ref{fig:2}\textbf{f,g} displays frequency and phonon lifetime of the $E_{2g}$ phonon as a function of temperature. The mode hardening can be modeled by characteristic anharmonic effects \cite{Balkanski1983prb} where $\omega (T) = \omega_0 + A( 1 + \frac{2}{e^x -1})+B(1+\frac{3}{e^y-1}+ \frac{3}{(e^y-1)^2})$ with $x= \hbar \omega_0 /2 k_B T$, $y=\hbar \omega_0 /3 k_B T $. Here, $\omega_0$ is the characteristic frequency at base temperature, $A$ and $B$ are constants. 
There are no anomalies in the frequency vs. temperature curve near the CDW transition. 
 On the other hand, the phonon lifetime shows a pronounced kink at $T_{\rm CDW}$, rising sharply toward a limiting low temperature value of 2.5 ps.  We carried out a similar analysis of the lower frequency phonons in ScV$_6$Sn$_6$ as well. No additional peak splitting or unusual phonon softening was identified within our sensitivity, although we emphasize that ScV$_6$Sn$_6$ is a very low brightness sample. 
Thus, the dramatic peak splitting of the 240 cm$^{-1}$ $A_{1g}$ symmetry vibrational  mode is key to understanding the CDW transition.

\begin{table}[b!]
      \caption{Subduction relations between the high-temperature $P6/mmm$ and low-temperature $R\bar{3}m$ structures. We only list the irreps of the high-symmetry structure that lead to Raman-active irreps in the low-temperature $R\bar{3}m$ structure. Raman active irreps are highlighted in bold. For the zone-center modes, we list both the space group irrep (for example $\Gamma_1^+$ and the corresponding point group irrep (such as $A_{1g}$.) Irrep labels $P_i$ refer to irreps at the $P \equiv (\frac{1}{3}, \frac{1}{3}, \frac{1}{3})$ point in reciprocal space for the $P6/mmm$ structure; our definition of these irreps is included in Supplementary table S3 (our notation matches that found on the Bilbao Crystallographic Server \cite{aroyo2006}).}
         \label{tab:correl}
    \centering
    \begin{tabular}{|ccc|}
        \multicolumn{1}{c}{ $P6/mmm$} & & \multicolumn{1}{c}{ $R\bar{3}m$} \\ \hline
        $\boldsymbol{A_{1g}/\Gamma_1^+}$ & $\rightarrow$ & $\boldsymbol{A_{1g}/\Gamma_1^+}$\\
        $B_{1g}/\Gamma_4^+$ & $\rightarrow$ & $\boldsymbol{A_{1g}/\Gamma_1^+}$\\
         $\boldsymbol{E_{1g}/\Gamma_6^+}$ & $\rightarrow$ & $\boldsymbol{E_g/\Gamma_3^+}$ \\
         $\boldsymbol{E_{2g}/\Gamma_5^+}$ & $\rightarrow$ & $\boldsymbol{E_g/\Gamma_3^+}$ \\
        $P_1$ & $\rightarrow$ & $\Lambda_1$ + $\boldsymbol{A_{1g}/\Gamma_1^+}$ + $A_{2u}/\Gamma_2^-$\\
        $P_3$ & $\rightarrow$ & $\Lambda_3$ + $E_{u}/\Gamma_3^-$+ $\boldsymbol{E_{g}/\Gamma_3^+}$ \\
        \hline
       \end{tabular}
\end{table}


 \subsection*{Phonon mixing in the low temperature phase}

In order to better understand the multiplet structure of the 240 cm$^{-1}$ phonon, we performed a symmetry analysis of the phonon modes. \cite{aroyo2006} In particular, we solved the \textit{subduction} problem which relates the irreps of the high-temperature $P6/mmm$ structure and the low-temperature $R\bar{3}m$ structure [Table \ref{tab:correl}]. This is a powerful technique that unravels a set of perturbed modes in terms of unperturbed vibrational modes.\cite{Long2000,Zhu2002} As expected, $A_{1g}$ irreps in $P6/mmm$, which exhibit the full crystal symmetry of the $P6/mmm$ structure, also exhibit the full crystal symmetry of the subgroup $R\bar{3}m$, and thus map to $A_{1g}$ in $R\bar{3}m$ as well. This irrep is Raman-active in both structures, but is only one-dimensional - the single $A_{1g}$ peak at 240 cm$^{-1}$ in the high-temperature structure \emph{cannot}, on its own, account for all five new modes in the $R\bar{3}m$ structure. Where, then, do the new modes come from?

\begin{figure*}[t!]
\centering
\includegraphics[width=\linewidth]{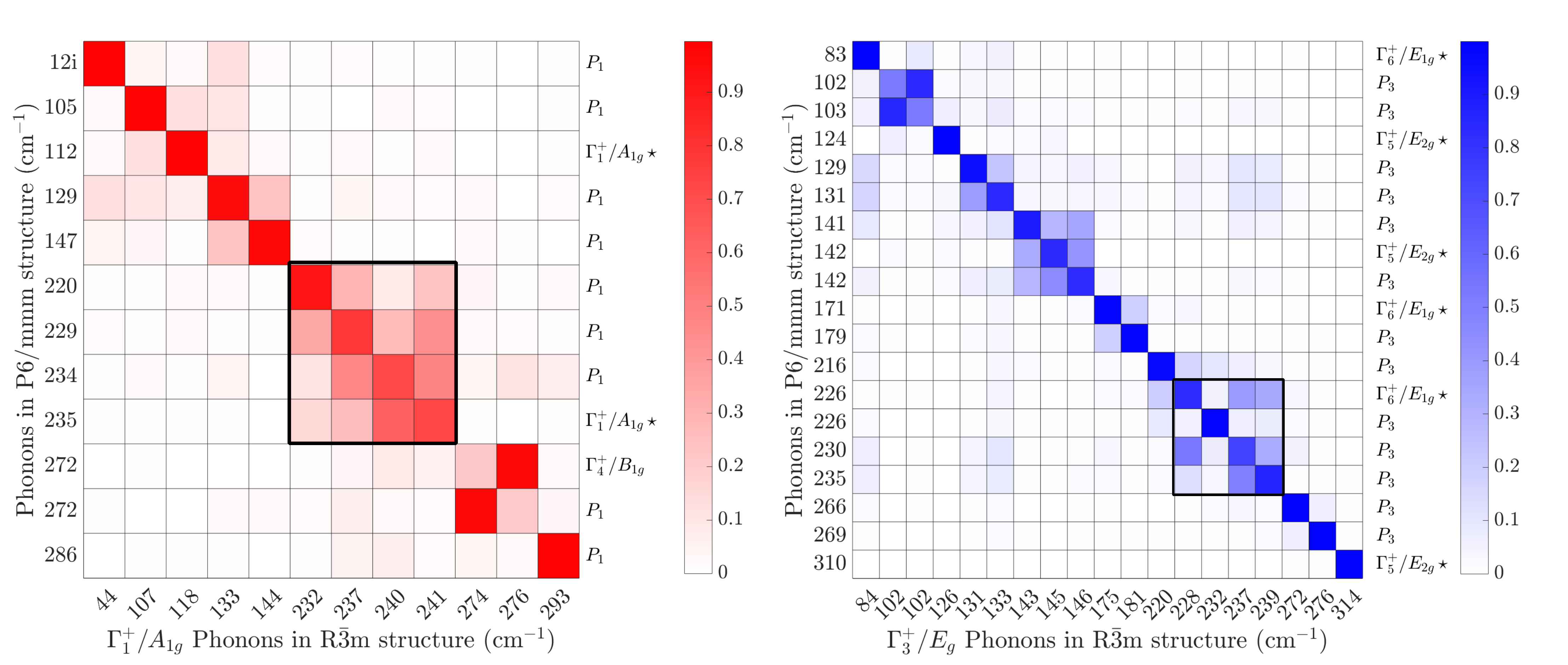}
\caption{Overlap of eigenvectors of the dynamical matrix between phonons in the $P6/mmm$ structure (vertical axis) and phonons in the $R\bar{3}m$ structure (horizontal axis). The color intensity of each box denotes the magnitude of the overlap, with a maximum possibly amplitude of unity. The lefthand panel corresponds to projections of $P6/mmm$ phonons transforming as $A_{1g}$, $B_{1g}$, and $P_1$ onto $R\bar{3}m$ phonons transforming as $A_{1g}$, while the righthand panel corresponds to projections of $P6/mmm$ phonons transforming as $E_{1g}$,  $E_{2g}$, and $P_1$ onto $R\bar{3}m$ phonons transforming as $E_{g}$. Taken together, these two panels include all Raman-active phonon modes in the $R\bar{3}m$ structure, as well as all $P6/mmm$ phonon modes related to them through subduction [Table \ref{tab:correl}]. Raman active modes in the $P6/mmm$ structure are denoted with a ``$\star$" character. 
 In both panels, the area of interest corresponding to phonons from 228 cm$^{-1}$ to 241 cm$^{-1}$ in the $R\bar{3}m$ structure are boxed. Note that the data as presented need not be symmetric.
}
\label{fig:overlaps}
\end{figure*}

The results in Table \ref{tab:correl} show that new Raman active modes can appear in $R\bar{3}m$ from the $B_{1g}$ modes of $P6/mmm$ at the zone center, as well as $P_1$ and $P_3$ modes which fold in to $\Gamma$ from the $P \equiv (\frac{1}{3}, \frac{1}{3}, \frac{1}{3})$ point on the Brillouin zone boundary in reciprocal space (defined in Supplementary table S3). In order to gain a quantitative understanding of the origin of each mode, we perform an analysis where the eigenvectors of the dynamical matrix associated with each phonon mode in the $R\bar{3}m$ structure are projected onto the eigenvectors of the dynamical matrix in the $P6/mmm$ structure as described in the Methods section. While all symmetry-allowed $P6/mmm$ phonon modes are allowed to mix in the formation of the $R\bar{3}m$ phonon modes (for example, a given $R\bar{3}m$ $A_{1g}$ mode could exhibit significant contributions from multiple $P6/mmm$ $A_{1g}$, $B_{1g}$, and $P_1$ modes simultaneously), this technique allows us to understand the relative magnitudes of those contributions. The results shown in Fig. \ref{fig:overlaps} show eight Raman-active phonon modes in the range of 228 cm$^{-1}$ to 241 cm$^{-1}$ in the $R\bar{3}m$ structure, which exhibit significant overlap with $A_{1g}$, $E_{1g}$, $P_1$, and $P_3$ phonon modes in the range of 220 cm$^{-1}$ to 235 cm$^{-1}$ in the $P6/mmm$ structure. Thus, the five clear peaks that result from the apparent splitting of the $A_{1g}$ peak at 240 cm$^{-1}$ observed in experiment are likely a subset of these eight modes, and therefore originate not just from the $A_{1g}$ peak at 240 cm$^{-1}$, but from $E_{1g}$ modes as well as modes at the $P$ point in reciprocal space which fold in to the $\Gamma$ point and become Raman-active in the $R\bar{3}m$ structure.

\subsection*{Pressure destabilizes the CDW transition}

\begin{figure*}[tbh]
\centering
\includegraphics[width=\linewidth]{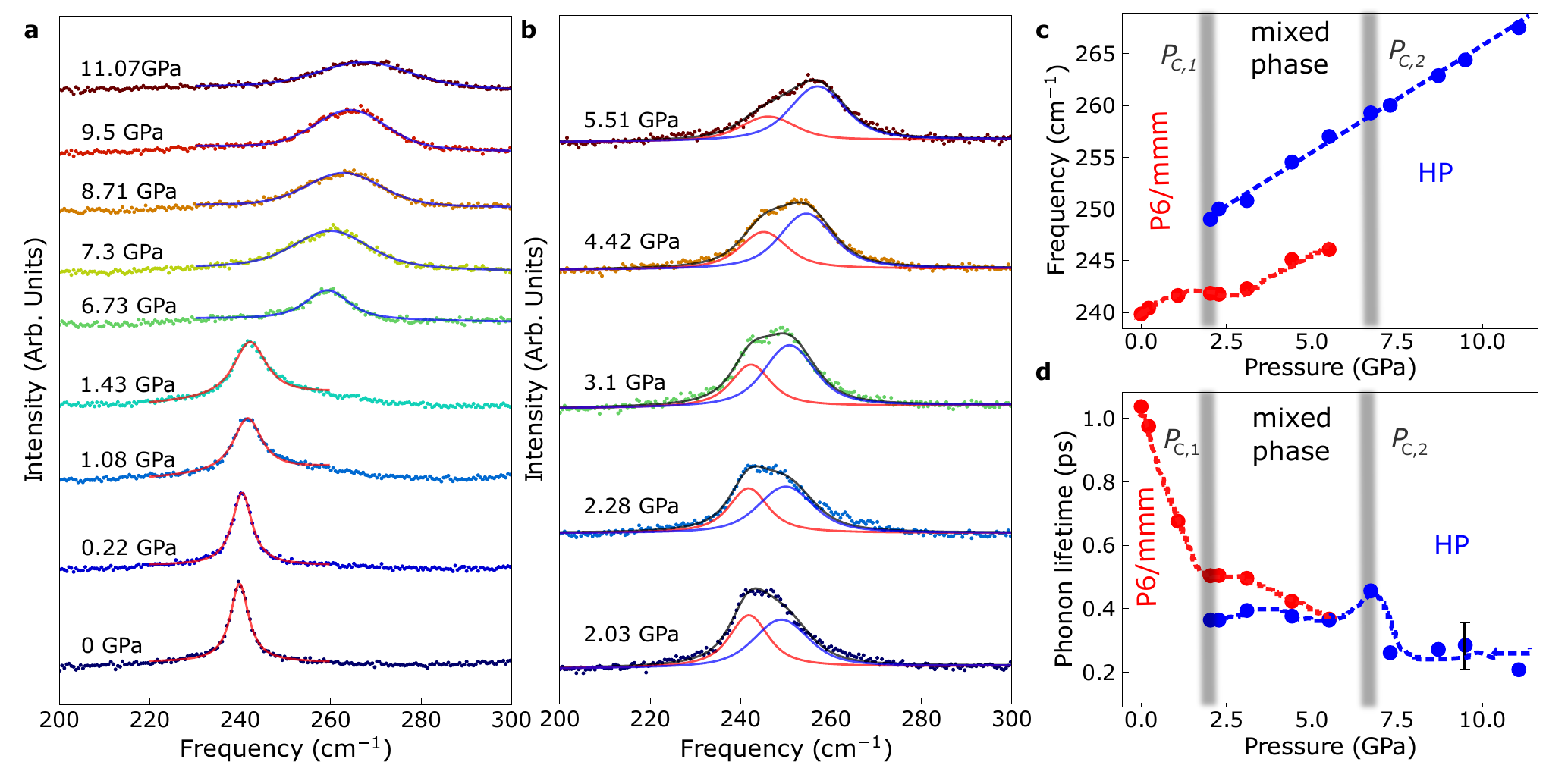}
\caption{\textbf{Pressure-driven structural phase transition in ScV$_6$Sn$_6$ at room temperature.} \textbf{a, b} Raman scattering response as a function of pressure.  The dotted lines are experimental data, and the solid lines correspond to fitted curves. \textbf{c, d} Raman shift and phonon lifetime as a function of pressure for the 240 cm$^{-1}$ phonon. Two critical pressures ($P_{\rm C,1}$ = 2 GPa and $P_{\rm C,2}$ = 6 GPa) separate the $P6/mmm$ phase, the mixed phase, and the high pressure (HP) phase. The error bars in panel \textbf{c} are on the order of the symbol size; a characteristic error bar is indicated in panel \textbf{d}.}
\label{fig:4}
\end{figure*}

Figure \ref{fig:4} summarizes the Raman scattering response of ScV$_6$Sn$_6$ under pressure at room temperature. We focus on the behavior of the $A_{1g}$ symmetry phonon near 240 cm$^{-1}$ because (i) it is most strongly connected to the CDW state and (ii) all other modes show only simple hardening under compression - at least within our sensitivity. As indicated in Fig. \ref{fig:4}\textbf{a,b}, this peak can be modeled with a single oscillator below 2 GPa and also above approximately 6 GPa. At intermediate pressures, the spectra of ScV$_6$Sn$_6$ are best fit with two Voigt oscillators indicating the presence of a mixed phase.  This type of mixed or two-phase regime is typical for a first-order pressure-driven transition. \cite{Musfeldt2023} In this case, the mixed phase consists of $P6/mmm$ + a new high pressure phase. 
Based on this behavior, we define two critical pressures: $P_{\rm C,1}$ = 2 GPa and $P_{\rm C,2}$ = 6 GPa.

Figure \ref{fig:4}\textbf{c,d} displays frequency vs. pressure and phonon lifetime trends for ScV$_6$Sn$_6$. 
We find that the concentration of $P6/mmm$ (as measured by the relative strength of the low frequency oscillator) diminishes with increasing compression in the mixed phase region whereas that of the new high pressure phase (measured by the relative strength of the second peak) increases until it dominates above $P_{\rm C,2}$. 
The high pressure phase of ScV$_6$Sn$_6$ is very similar to but not identical to $P6/mmm$.  The single $A_{1g}$-like phonon mode emerging above $P_{\rm C,2}$ has a higher frequency than that in the $P6/mmm$ ground state, pointing to a  stronger force constant 
as well as a  volume reduction in the high pressure phase which may have interesting connections to the inter-layer coupling in this class of kagom\'e metals.
At the same time, phonon lifetime drops from 1 ps at ambient pressure (lower than most traditional semiconductors and chalcogenides) to approximately 0.2 ps over this range. The decreased lifetime is due to significantly increased scattering events under compression. \cite{sun2013spectroscopic}  We mention in passing that we could not stabilize the high pressure phase in our first principles simulations under pressure. This may be due to complications from the residual density wave state recently reported at room temperature or some other shortcoming of density functional theory.\cite{cao2023competing, korshunov2023softening}

Prior transport studies reveal that the structure of ScV$_6$Sn$_6$ starts to change near 2 GPa due to destabilization and disappearance of the CDW. \cite{zhang2022destabilization} It is  tempting to claim that this destabilization corresponds to $P_{\rm C,1}$, but we must remember that our measurements are performed at room temperature - where the CDW is not fully developed. The recently reported short range CDW provides a way forward. This ``residual CDW" resides in the high temperature phase of ScV$_6$Sn$_6$ and hosts a  (1/3,1/3,1/2) propagation vector. \cite{cao2023competing, korshunov2023softening} We hypothesize that pressure quenches short range CDW correlations in this system as well, although it appears to do so in a first-order mixed phase manner between $P_{\rm C,1}$ and $P_{\rm C,2}$. 

There are not too many parallels between the pressure induced transition we observe here in ScV$_6$Sn$_6$ and those observed in the single-layer vanadate kagom\'es like $A$V$_3$Sb$_5$, because the latter takes place at low temperature and between two different long-range-ordered CDW phases and the high symmetry phase, whereas in ScV$_6$Sn$_6$ this transition takes place at room temperature, well above the long-range CDW order.\cite{zhang2021pressure, gupta2022two, ritz2022impact} 
A possibility that explains the disappearance of CDW at high pressures is that the system undergoes a volume-collapse transition similar to those observed in the ThCr$_2$Si$_2$ family of compounds,  which involves the formation of Sn--Sn covalent bonds.\cite{hoffmann2002making, foroozani2014hydrostatic} Such transitions are also observed in various iron-pnictide superconductors, and couple strongly to superconductivity because of their effect on the Fermi surface\cite{kreyssig2008, saha2012}. Therefore, such a transition is likely to suppress CDW as well. 

Finally, we note that similar measurements on the Lu analog reveal no evidence at all for this type of pressure-induced destabilization of the short range CDW [Supplementary figure 3]. The A$_{1g}$ symmetry mode near 240 cm$^{-1}$ hardens systematically under compression in line with the fact that LuV$_6$Sn$_6$ does not host a CDW. \cite{lee2022anisotropic} These structure-property relations are 
nicely unified by $R$ site size arguments [Inset, supplementary figure 1\textbf{d}]  which reveal that both chemical and physical pressure drive similar trends in CDW stability.

\subsection*{Unraveling CDW stability in the $R$V$_6$Sn$_6$ family of materials ($R$ = Sc, Y, Lu)}

Traditionally,  CDW behavior is related to a periodic lattice modulation due to Fermi surface nesting and electron-phonon interactions.\cite{zhu2015classification,gruner1985charge,roy2021quasi,brouet2004fermi,weber2011electron,jurczek1986charge} 
Unraveling the complete mechanism requires a systematic investigation of the electronic structure 
as well as a full exploration of the vibrational properties. In this family of materials, most of the focus so far has been on electronic properties from the point of view of electron-phonon coupling and Fermi surface nesting.\cite{lee2023,cheng2023nanoscale,kang2023emergence,hu2023phonon,tuniz2023dynamics} 
There has been significantly less effort to uncover the vibrational contribution to the development of the density wave state. This is because 
metallic character in both high and low temperature phases of ScV$_6$Sn$_6$  prevents us from revealing the behavior of the infrared-active phonons due to  screening by the Drude peak. The Raman scattering response is, however, still accessible, and combined with different external stimuli and complementary lattice dynamics calculations, a remarkable picture of phonon mixing and density wave stability is beginning to emerge.

In this work, we traced how phonons mix in the CDW state of ScV$_6$Sn$_6$, tested these ideas with symmetry arguments based upon our lattice dynamics calculations, and compared our findings to behavior in the Y and Lu analogs. Overall, we demonstrate that low temperature stabilizes the CDW whereas both physical and chemical pressures destroy it. This suggests that the coupling interactions are on the ``knife's edge", easily manipulated by any external stimuli, and more sensitive than the similar CDW orders in the monolayer AV$_3$Sb$_5$ vanadates. 
Our analysis reveals that the multiplet-like behavior of the 240 cm$^{-1}$ phonon in the CDW state of ScV$_6$Sn$_6$ is the result of its mixing with other phonons. We emphasize that this splitting is not due to symmetry breaking of a single mode in the conventional sense. Instead, it is connected with both $\Gamma$ and $P$ point modes that are folded to the zone center and become Raman active, indicating that the CDW state is influenced by a variety of phonon modes and highlighting the intricate, phonon-assisted nature of the CDW state. 
Since the phonon mode near 240 cm$^{-1}$ also serves to connect the short-range and long-range CDWs, it provides a sensitive, microscopic indicator of CDW stability under different tuning parameters. Finally, we point out that while prior phonon dispersion calculations emphasize the role of unstable Sn and Sc modes in the CDW state, \cite{tan2023abundant} this and other work \cite{arachchige2022charge,cheng2023nanoscale,tuniz2023dynamics} is revealing that the V centers are not just spectators. 
Rather, the out-of-plane motion of V in the double kagom\'e layer is just as significant as the Sc and Sn center motion. 
Recent STM work reaches a similar conclusion regarding the importance of the V centers. \cite{cheng2023nanoscale}

\section*{Methods}

\paragraph*{\bf Crystal growth and diamond anvil cell loading:} High quality $R$V$_6$Sn$_6$ ($R$= Sc, Lu, Y) crystals were grown from a Sn-rich melt method with a composition ratio of $R$:V:Sn = 1:6:60 as described in Ref. \cite{arachchige2022charge}. They are hexagonal blocks with clear $ab$ surfaces. For the high pressure Raman scattering measurements, a small single crystal was loaded into a symmetric diamond anvil cell suitable for work in the 0 - 13 GPa range. The cell is equipped with low fluorescence diamonds with 400 $\mu$m culets. We also employed a stainless steel gasket with a 100 $\mu$m hole, KBr as the pressure medium, and an annealed ruby ball for pressure determination via fluorescence. \cite{Mao1976}

\paragraph*{\bf Raman spectroscopy:} Raman scattering measurements were performed in the back scattering geometry using a Horiba LabRAM HR Evolution spectrometer equipped with a 532 nm (green) laser, a 50$\times$ microscope objective, 1800 line/mm gratings, and a liquid-nitrogen-cooled charge-coupled device detector.  To minimize heating and maximize signal intensity for this low brightness sample, power was controlled below 7.7 mW, and the laser was slightly defocused. Each spectrum was integrated for 200 s and averaged four times. Variable temperature work was carried out with a low-profile open-flow cryostat, and high pressure measurements employed a diamond anvil cell as described above. Standard peak fitting techniques were employed as appropriate.

\paragraph*{\bf Lattice dynamics calculations:}

All density functional theory (DFT) calculations were performed with Projector Augmented Waves (PAW) as implemented in the Vienna Ab initio simulation package (VASP) version 5.4.4 \cite{kresse1993ab,kresse1996efficiency,kresse1996efficient} using the PBEsol exchange-correlation functional
for valence configurations of Sc, V, and Sn corresponding to 3\emph{s}$^2$3\emph{p}$^6$4\emph{s}$^1$3\emph{d}$^2$, 3\emph{s}$^2$3\emph{p}$^6$4\emph{s}$^1$3\emph{d}$^4$, and 5\emph{s}$^2$4\emph{d}$^{10}$5\emph{p}$^2$, respectively. Unless otherwise mentioned, experimentally determined lattice parameters of $a=5.475$ {\AA} and $c=9.177$ {\AA} for the $P6/mmm$ structure and $a=9.456$ {\AA} and $c=27.412$ {\AA} for the conventional $R\bar{3}m$ structure were used for all calculations  
Internal degrees of freedom were relaxed, with forces converged to within 0.001 {eV/\AA} using a plane wave cutoff energy of 400 eV, combined with a $\Gamma$-centered Monkhorst-Pack k-point mesh of 20$\times$20$\times$10 in the $P6/mmm$ structure, as well as a Gaussian smearing parameter of 10 meV. Calculations in the $R\bar{3}m$ structure were carried out in a primitive, three formula-unit cell commensurate with the wavevector $P=(\frac{1}{3}, \frac{1}{3}, -\frac{1}{3})$, which corresponds to a unit cell with basis vectors $(\frac{2}{3},\frac{1}{3},\frac{1}{3}),(-\frac{1}{3},\frac{1}{3},\frac{1}{3}),(-\frac{1}{3},-\frac{2}{3},\frac{1}{3})$, where the indices $(a,b,c)$ correspond to the lattice vectors of the conventional $R\bar{3}m$ nine formula unit cell. All computational parameters used for this structure were the same as for the $P6/mmm$ structure, except with a $\Gamma$-centered Monkhorst-Pack k-point mesh of 10$\times$10$\times$10. To calculate phonon frequencies and their associated distortions, we constructed the dynamical matrix in a basis of symmetry adapted modes, which bring the dynamical matrix into block diagonal form, where each block corresponds to a single irreducible representation of the space group. These symmetry-adapted linear combinations of atomic displacements were found using the ISOTROPY software suite. \cite{stokes2022} 

\paragraph*{\bf Phonon overlap:}
After calculating the eigenvectors of the dynamical matrix associated with each phonon mode, we then computed the relationship between eigenvectors $\hat{e}_i^{I\alpha}$ in the $P6/mmm$ structure (where $I$ is an irrep, $\alpha$ the mode index within that irrep, and $i$ the ionic degree of freedom in cartesian coordinates), and the eigenvectors $\hat{g}_i^{I\alpha}$ in the $R\bar{3}m$ structure. First, the $\hat{e}_i^{I\alpha}$ were expressed in a three formula unit basis commensurate with the $R\bar{3}m$ cell, then renormalized. The overlap for two phonons is then defined as the projection of $\hat{g}_i^{I\alpha}$ onto $\hat{e}_i^{I\alpha}$, 

\begin{equation}
 O(J,\beta,I,\alpha)=\sqrt{\left(\hat{g}_i^{J\beta}\hat{e}_i^{I\alpha}\right)^2}.
\end{equation}

In the lefthand panel of Fig. \ref{fig:overlaps}, $J=A_{1g}$, and $I\in A_{1g}, B_{1g}, P_1$, as determined by the solution to the subduction proble described in Table \ref{tab:correl}. In the righthand panel, $J=E_{g}$, and $I\in E_{1g}, E_{2g}, P_3$. In this second case, the $E_{1g}$ and $E_{2g}$ phonons are doubly degenerate, and the $P_3$ phonons quadruply degenerate. In order to compress the results of the table, instead of reporting a cell for each individual degenerate mode, our figure includes a single cell with amplitude defined as $\tilde{O}$, where

\begin{equation}
 \tilde{O}(J,\beta,I,\alpha)\equiv\frac{1}{M_{\beta}}\sum_{\beta_l}\sqrt{\sum_{\alpha_m} O(J,\beta_l,I,\alpha_m)^2}.
\end{equation}
\noindent
Here, $l$ and $m$ run over the degenerate indices of $\beta$ and $\alpha$ respectively, and ${M_{\beta}}$ is the multiplicity of the $\beta$ irrep being projected.
Similar approaches have been employed in other materials.\cite{Long2000, Zhu2002}

\paragraph*{\bf Group Theory:}

The symmetry-adapted linear combinations of atomic displacements used for the phonon calculations were found using the ISOTROPY software suite.\cite{stokes2022} 
The subduction analysis to find the connections between irreps through the group-subgroup transition was performed using the CORREL application hosted by the Bilbao crystallographic server.\cite{aroyo2006}


\section*{Data availability}

Data are available from the corresponding authors upon reasonable request.

\section*{Acknowledgements}

Work at Tennessee (YG, KAS, ALB, and JLM) is supported by Physical Behavior of Materials, Basic Energy Sciences, U.S. Department of Energy (Contract number DE-SC00023144).  DM and WRM acknowledges funding from the Gordon and Betty Moore Foundation’s EPiQS Initiative, Grant GBMF9069. Work at the University of Minnesota (ETR and TB) was supported by NSF CAREER grant DMR-2046020.

\section*{Author contributions}

YG and JLM designed the study. WRM, SM, and RPM grew the crystals with guidance from DGM. ALB carried out feasibility measurements while YG performed the variable temperature and high pressure Raman scattering spectroscopy. ER performed the theoretical calculations and discussed the results with TB. YG and KAS analysed the spectral data with guidance from JLM. YG, ER, and JLM wrote the manuscript. All authors commented on the text.

\section*{Competing interests}

The authors declare no competing financial or non-financial interests.

\nocite{*}

\bibliography{ref}

\end{document}


\preprint{APS/123-QED}

\title{Supplemental information: Origin and stability of the charge density wave in ScV$_6$Sn$_6$}

\author{Yanhong Gu}
\affiliation{Department of Chemistry, University of Tennessee, Knoxville, Tennessee 37996, USA}

\author{Ethan Ritz}
\affiliation{Department of Chemical Engineering and Materials Science,
University of Minnesota, Minneapolis, Minnesota 55455, USA}

\author{William R. Meier}
\affiliation{Materials Science and Engineering Department, University of Tennessee Knoxville, Knoxville, Tennessee 37996, USA}
\author{Avery Blockmon}
\affiliation{Department of Chemistry, University of Tennessee, Knoxville, Tennessee 37996, USA}
\author{Kevin Smith}
\affiliation{Department of Chemistry, University of Tennessee, Knoxville, Tennessee 37996, USA}

\author{Richa Pokharel Madhogaria}
\author{Shirin Mozaffari} 
\affiliation{Materials Science and Engineering Department, University of Tennessee Knoxville, Knoxville, Tennessee 37996, USA}
\author{David Mandrus}
\affiliation{Materials Science and Engineering Department, University of Tennessee Knoxville, Knoxville, Tennessee 37996, USA}
\affiliation{Department of Physics and Astronomy, University of Tennessee Knoxville, Knoxville, Tennessee 37996, USA}
\affiliation{Materials Science and Technology Division, Oak Ridge National Laboratory, Oak Ridge, Tennessee 37831, USA}

\author{Turan Birol}
\affiliation{Department of Chemical Engineering and Materials Science,
University of Minnesota, Minneapolis, Minnesota 55455, USA}

\author{Janice L. Musfeldt}
 \email{musfeldt@tennessee.edu}
\affiliation{Department of Chemistry, University of Tennessee, Knoxville, Tennessee 37996, USA}
\affiliation{Department of Physics and Astronomy, University of Tennessee, Knoxville, Tennessee 37996, USA}
\date{\today}%

\maketitle

\tableofcontents

\section{SUPPLEMENTARY NOTE 1: Assignment of the Raman-active modes in ScV$_6$Sn$_6$}

We employed a number of different techniques to clarify the mode assignment of ScV$_6$Sn$_6$. These include a symmetry analysis, first-principles lattice dynamics calculations, measuring the rotation modes of air,  utilizing polarization control, and substituting rare-earth elements into the $R$ site. We also performed both temperature and pressure dependence of confirm vibrational mode assignments. 

We began by reviewing the low frequency features (below 200 cm$^{-1}$) of the Raman spectra. We measured Raman spectra on Si wafer to obtain the rotation modes of air\cite{ohno2021pure}, as there are almost no Raman features of Si in the range of 50 - 200 cm$^{-1}$, and also Si wafer has a low scattering rate in this range. Therefore, a Raman spectra with rotation modes of air can be extracted. We compared the Raman spectra of Si wafer and ScV$_6$Sn$_6$ in Supplementary Figure \ref{fig:1}\textbf{a}. By comparing the Raman spectra of the Si wafer and ScV$_6$Sn$_6$, two peaks marked by the orange arrows of ScV$_6$Sn$_6$ can be identified as $E_{1g}$ (near 87 cm $^{-1}$)  and $E_{2g}$ (near 141 cm $^{-1}$) directly below 200 cm$^{-1}$. And above 200 cm$^{-1}$, three theory predicted modes are assigned to the experimental peaks accordingly. 
Since the rotation modes of air are mainly determined by the optical path, they do not change when we cool the sample to low temperatures. While, the phonons of ScV$_6$Sn$_6$ sample will change their position with cooling. We compared the spectra at room temperature and at the base temperature in Supplementary Figure \ref{fig:1}\textbf{b}. Then, we found that peak marked with the orange star shifts to high frequency at $T$ = 5 K, which should correspond to the $E_{2g}$ mode. Still, we have two theory predicted modes that cannot be resolved directly. The peaks marked with orange triangles are potentially modes of ScV$_6$Sn$_6$, and we assign them based on the relative difference in frequency between the experimental results and calculations. But they are weak and hard to distinguish on their own.

The $E_{1g}$ mode near 240 cm$^{-1}$ is very weak and only shows a small shoulder in the spectra collected on $ac/bc$ plane. To confirm that the large peak near 240 cm$^{-1}$ is indeed the $A_{1g}$ mode, we combined polarizer and analyzer in the Raman measurement. The linearly polarized Raman spectra were collected on $ab$ plane of ScV$_6$Sn$_6$ single crystal. The incident green laser with wavelength 532 nm was normal to the $ab$ surface and the scattered beam was anti-paralle to the incident light. The polarization of incoming beam is controlled by a half wave plate. The experimental geometry is described by Porto’s notation\cite{ferraro2003introductory}. Here, $Z$ direction is along the $c$ axis, which is perpendicular to the $ab$ plane, and $X$ is along the $a$ axis of the sample. $Y$ is perpendicular to the $a$ axis. In Supplementary Figure \ref{fig:1}\textbf{c}, two different geometries are used for the measurement at room temperature, $Z(YY)\bar{Z}$ (the green solid line) and $Z(XY)\bar{Z}$(the red solid line). $Z$($\bar{Z}$) is the incoming(outgoing) light direction, $YY$ ($XY$) are the polarization directions of the incoming light(the first character) and  outgoing light(the second character). For example, $YY$ indicates that the analyzer is aligned parallel to the polarization axis of the incoming light, meaning that the incoming light polarization is parallel to the outgoing light polarization and perpendicular to the $a$ axis of the sample. The blue spheres in the plot represent the calculated phonons. Based on the polarization selection rules of $P6/mmm$, $A_{1g}$ and $E_{2g}$ are allowed in $Z(YY)\bar{Z}$ geometry and $E_{2g}$ is allowed in the $Z(XY)\bar{Z}$ geometry.  We focus on the modes near 240 cm$^{-1}$. The  240 cm$^{-1}$ peak disappears totally in the $Z(XY)\bar{Z}$ and appears in $Z(YY)\bar{Z}$ geometry, which confirms the strong peak near 240 cm$^{-1}$ consists primarily of the $A_{1g}$ phonon.

 \begin{figure}[h!]
\centering
\includegraphics[width=0.9\linewidth]{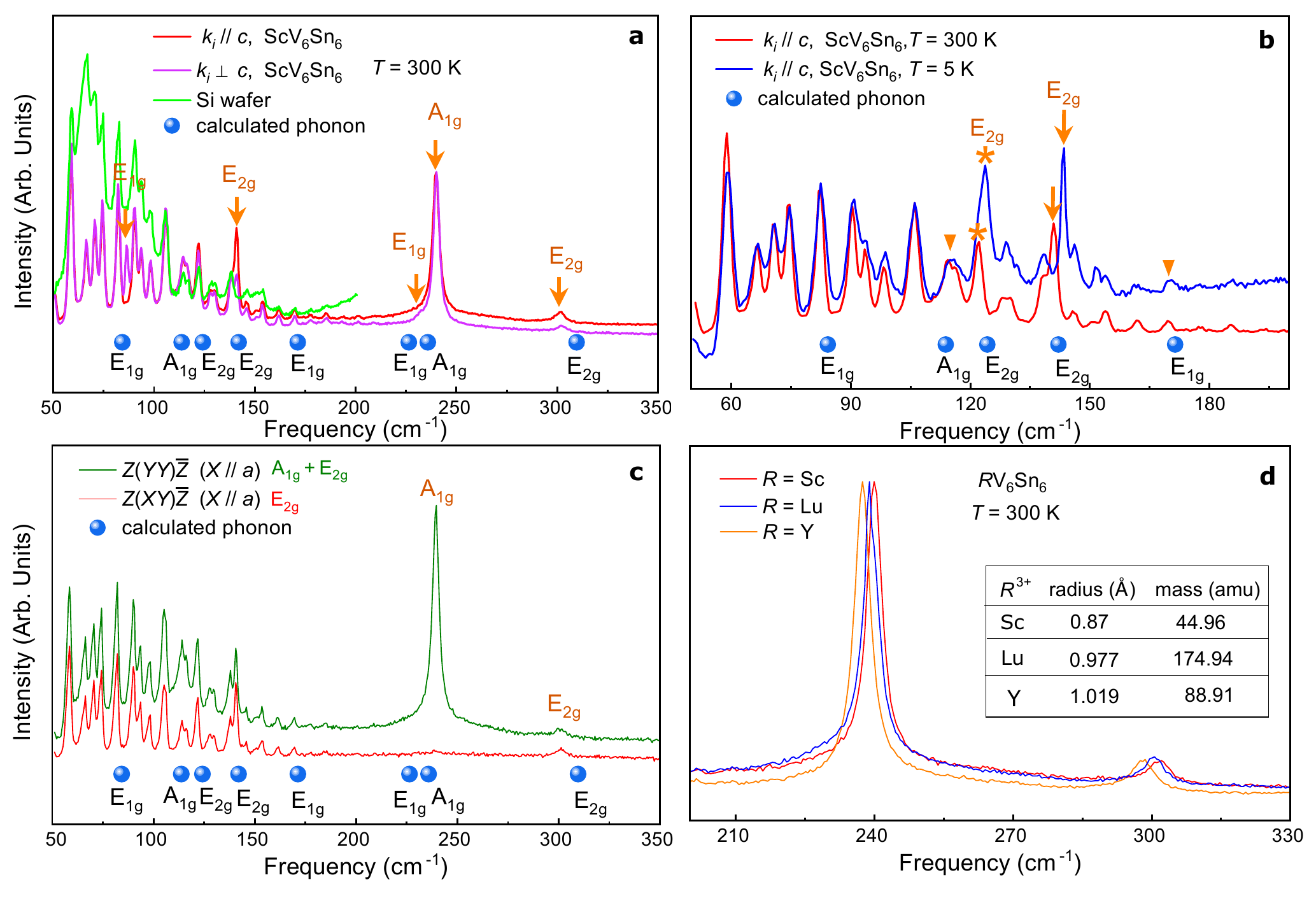}
\caption{\textbf{a} Comparison of Raman spectra at room temperature. Raman spectra of ScV$_6$Sn$_6$ are collected on the $ab$ plane (red solid line, the incident laser is parallel to the $c$ axis, $k_i \parallel c$) and $ac$/$bc$ plane (pink solid line, the incident laser is perpendicular to the $c$ axis, $k_i \perp c$).  Raman spectra of Si wafer is collected on the $ab$ plane (green solid line). The blue spheres are theory calculated Raman phonon frequencies. \textbf{b} Raman spectra of ScV$_6$Sn$_6$ collected at room temperature (red) and base temperature 5 K (blue). \textbf{c} Comparison of Raman spectra collected in $A_{1g}$ + $E_{2g}$ symmetry (green solid line) and $E_{2g}$ symmetry (red solid line) at room temperature. \textbf{d} 
 Spectra comparison of $R$V$_6$Sn$_6$ with $R$ = Sc (red), $R$ = Y (orange) and $R$ = Lu (blue). The spectra are collected on the $ab$ plane of the single crystal at room temperature. The inset summarizes the mass and radius of the $R$ center \cite{shannon1976revised}.}
\label{fig:1}
\end{figure}

 The Raman spectra comparison of ScV$_6$Sn$_6$, YV$_6$Sn$_6$, and LuScV$_6$Sn$_6$ is shown in Supplementary Figure \ref{fig:1}\textbf{d}. Three compounds show similar low frequency peaks. We focus on the peaks between 200 cm$^{-1}$ and 320 cm$^{-1}$. The frequency of peaks near 240 cm$^{-1}$ and 300 cm$^{-1}$ is highest
in ScV$_6$Sn$_6$, lowest in YV$_6$Sn$_6$, and LuV$_6$Sn$_6$ is in between. 
While their mass order is Lu, Y, and Sc, with Lu being the heaviest and Sc being the lightest element. Clearly, the peak shifts do not obey the isotope effect\cite{ferraro2003introductory}, i.e., the Raman frequency is proportional to the
$ \sqrt{k / \mu}$ ( where $k$ is the force constant and $\mu$ is the effective mass). This suggests that the local environment between the rare earth elements and V, Sn atoms is different and plays a significant role in these Raman modes. On the other side, the order of $R$ site radius is Y, Lu, and Sc, with Y having the largest radius and Sc having the smallest[Inset table in Supplementary  Figure 1\textbf{d}]. There appears to be a correlation between the radius of the rare earth element and the Raman frequency shifts. Comparison of the $R$ site Raman spectra suggests that the modes near 240 cm$^{-1}$ and 300 cm$^{-1}$ are highly sensitive to the $R$ site element. Here, we present a summary of the symmetries and the displacement patterns of Raman-active phonon modes in ScV$_6$Sn$_6$ in Supplementary Table 1.

\begin{longtable}{ | c |  c| c | c| c |c |}

\caption{Summary of Raman-active phonon mode symmetries, experimental and theoretical frequencies, and their corresponding displacement patterns of ScV$_6$Sn$_6$ computed with $P6/mmm$ symmetry.
}\label{tbl}
\\
\hline
\multirow{8} *{Symmetry} & \multirow{8} *{\makecell{ Experiment\\(cm$^{-1}$)}}  &  \multirow{8} *{\makecell{Theory \\(cm$^{-1}$)}}  & \multicolumn{3}{c|}{\multirow{2} *{Displacement Patterns}} \\
    &   &    & \multicolumn{3}{c|}{}  \\
  \cline{4-6}
    &   &   & {\makecell{$a$ axis view \\ \raisebox{-\totalheight}{\includegraphics[width=0.06\linewidth] {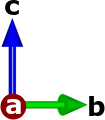} } } }& {\makecell{$b$ axis view \\ \raisebox{-\totalheight}{\includegraphics[width=0.075\linewidth] {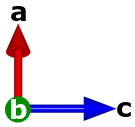} } } } & {\makecell{$c$ axis view \\ \raisebox{-\totalheight}{\includegraphics[width=0.09\linewidth] {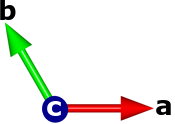} } } } \\
    &   &   &   &   &     \\
    \hline
  \multirow{-10} *{$E_{1g}$} & \multirow{-10} *{86.6} & \multirow{-10} *{83} &  \raisebox{0pt}{\includegraphics[width=0.2\linewidth] {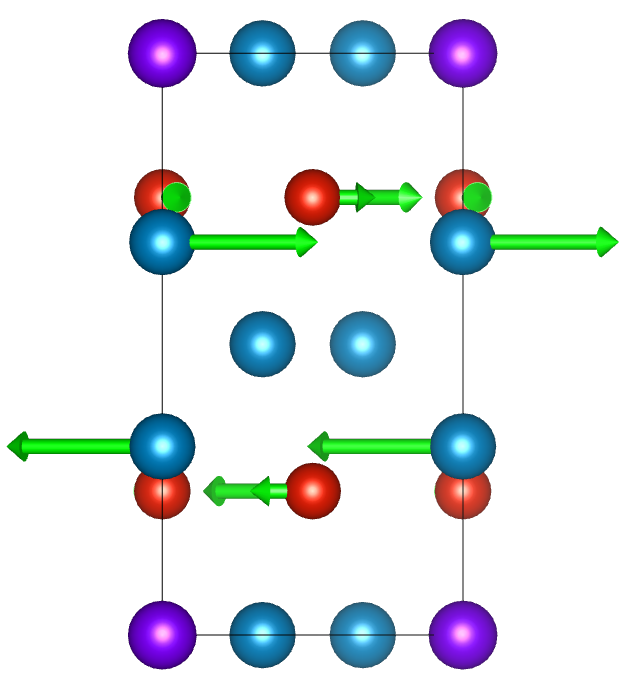} }& \raisebox{8pt} {\includegraphics[width=0.2\linewidth] {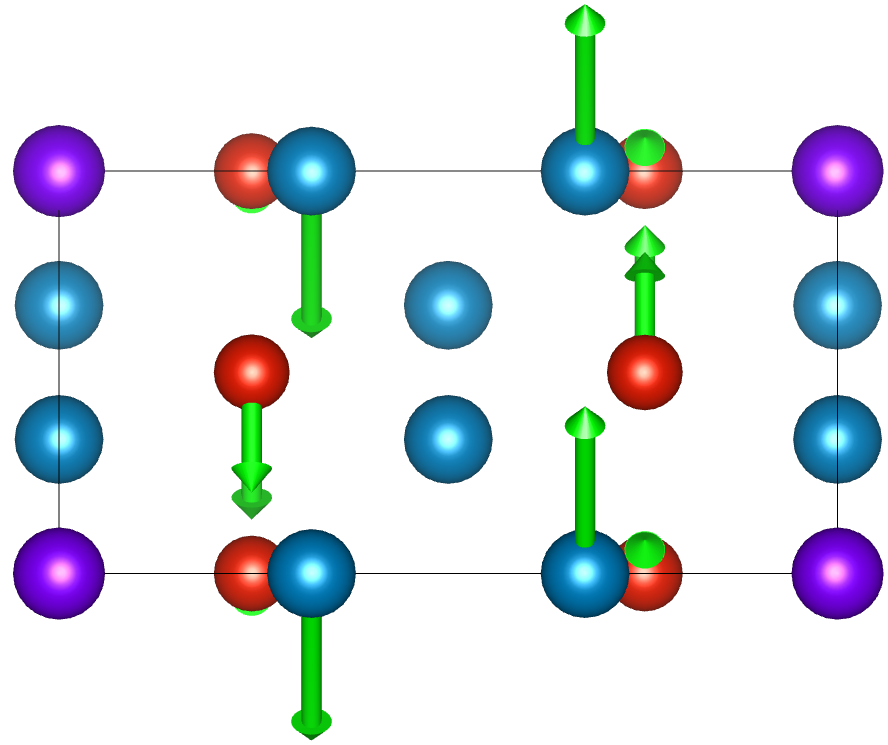}} & \raisebox{8pt} {\includegraphics[width=0.2\linewidth] {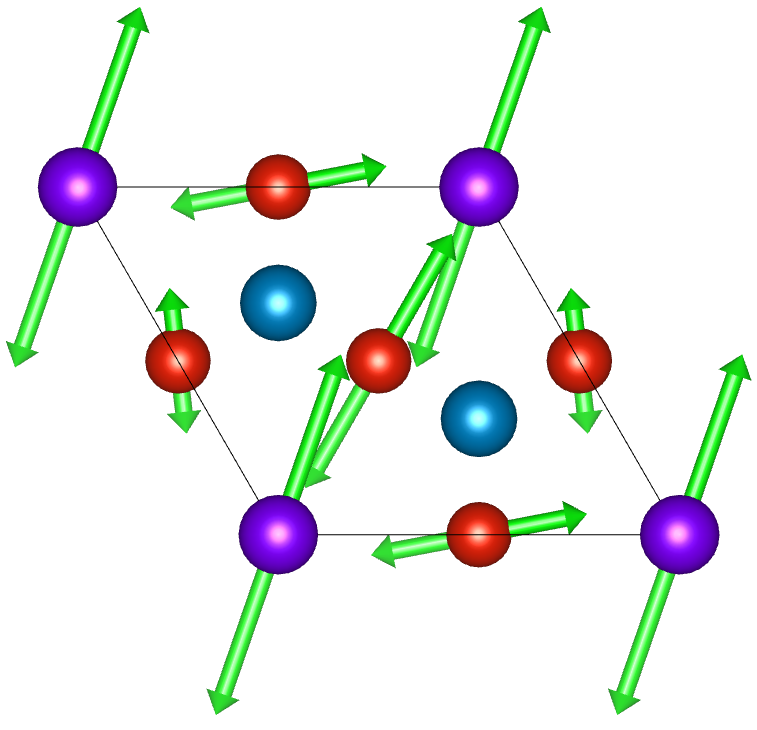}}\\
\hline
\multirow{-4} *{$A_{1g}$} & \multirow{-4} *{114}  & \multirow{-4} *{112} &   \raisebox{-38pt}{\includegraphics[width=0.12\linewidth] {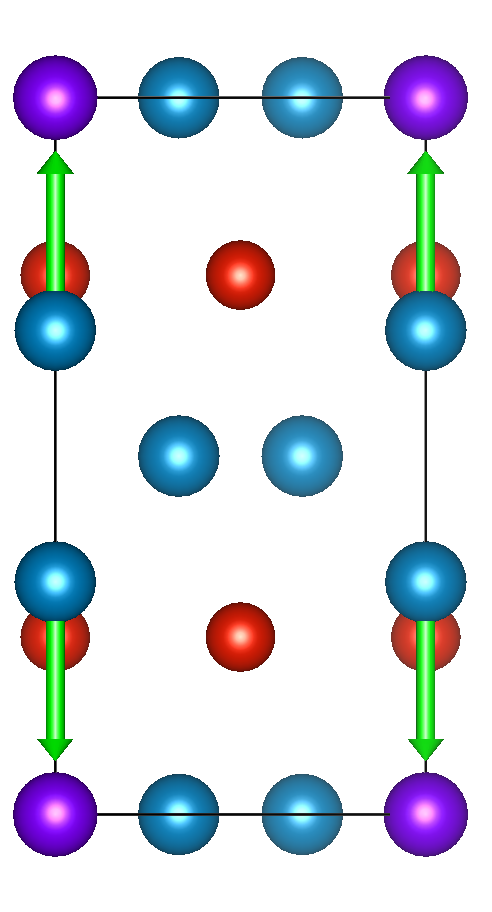}} &  \raisebox{-10pt} {\includegraphics[width=0.2\linewidth] {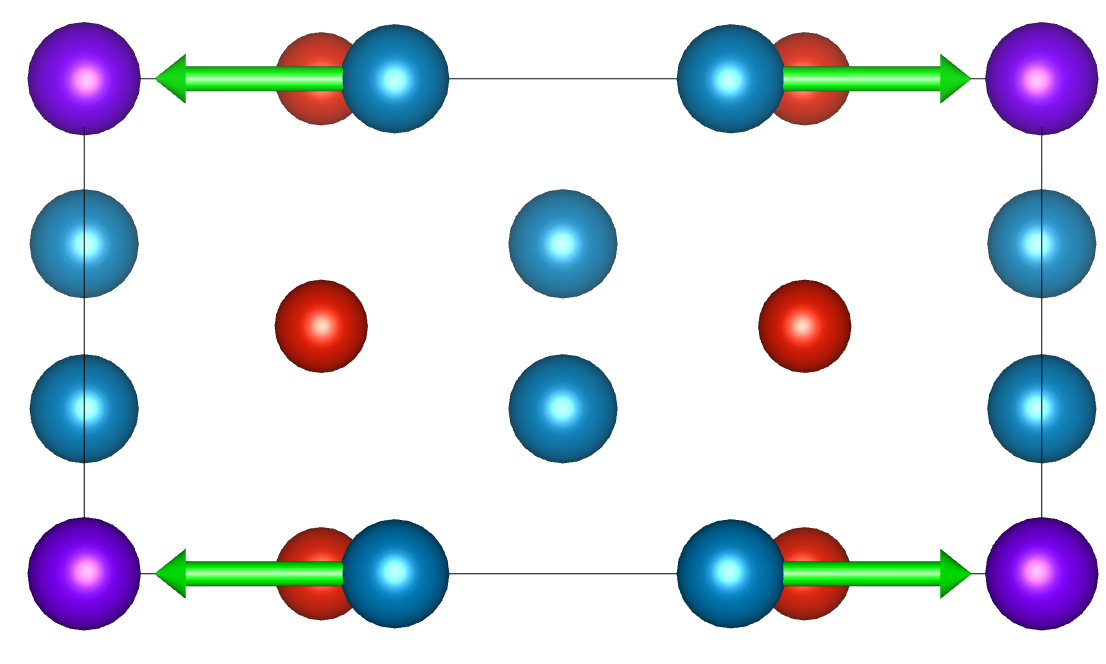} }&\raisebox{-10pt} {\includegraphics[width=0.2\linewidth] {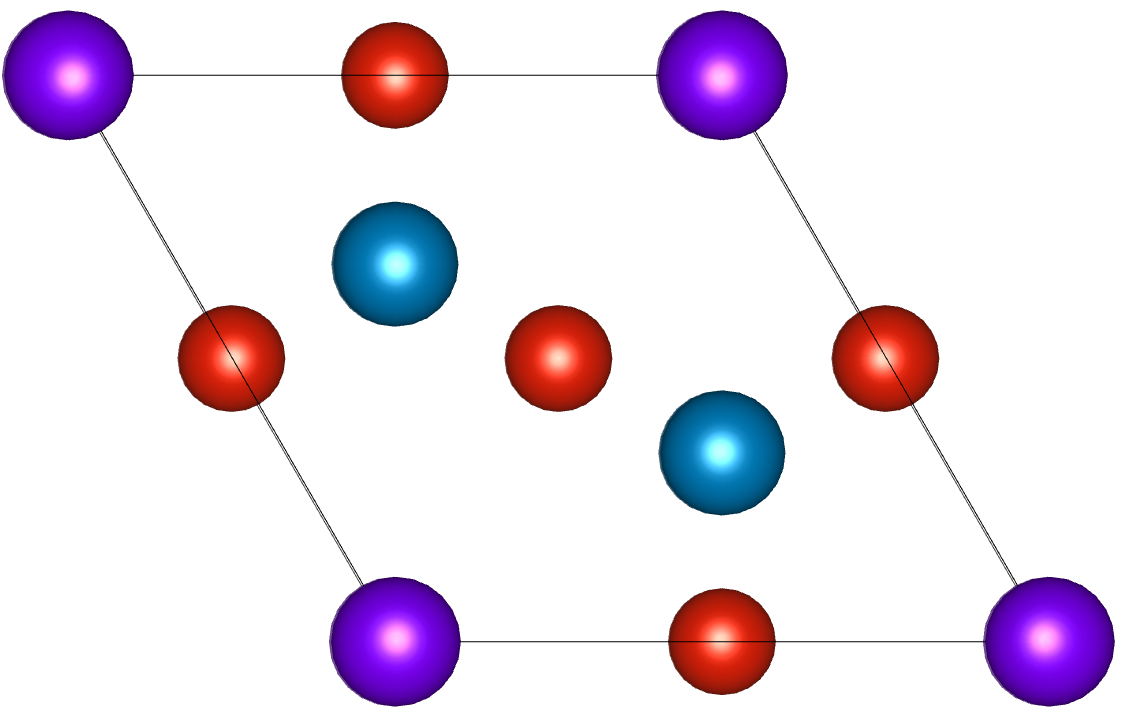}}\\
 \hline

   \multirow{-12} *{$E_{2g}$} & \multirow{-12} *{122.2} & \multirow{-12} *{124} &  \includegraphics[width=0.12\linewidth] {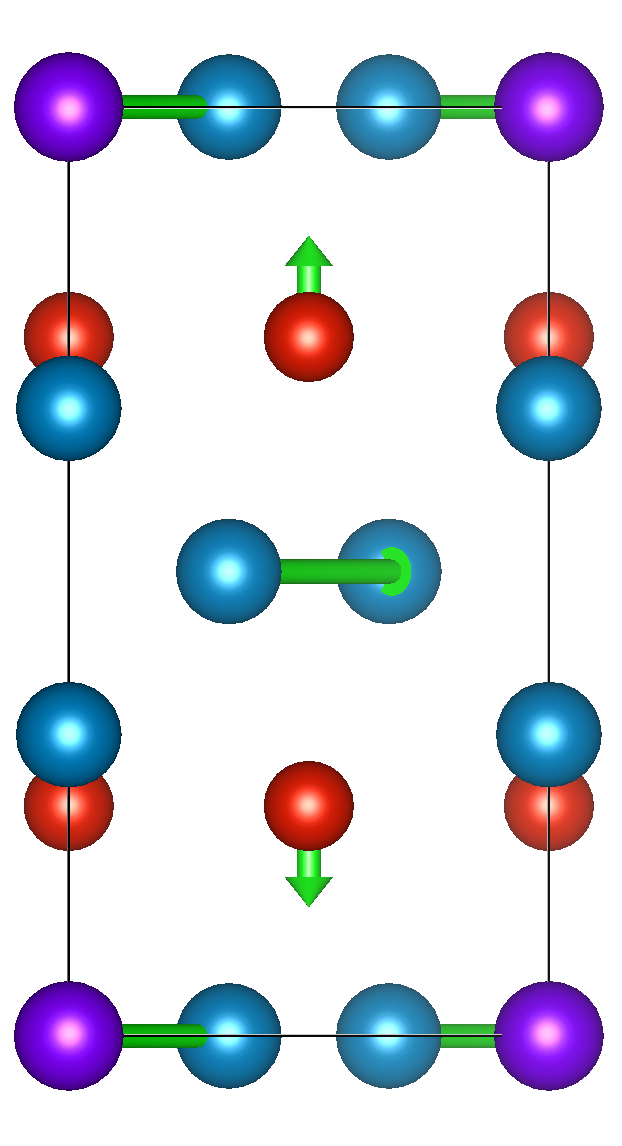} & \raisebox{18pt} {\includegraphics[width=0.2\linewidth] {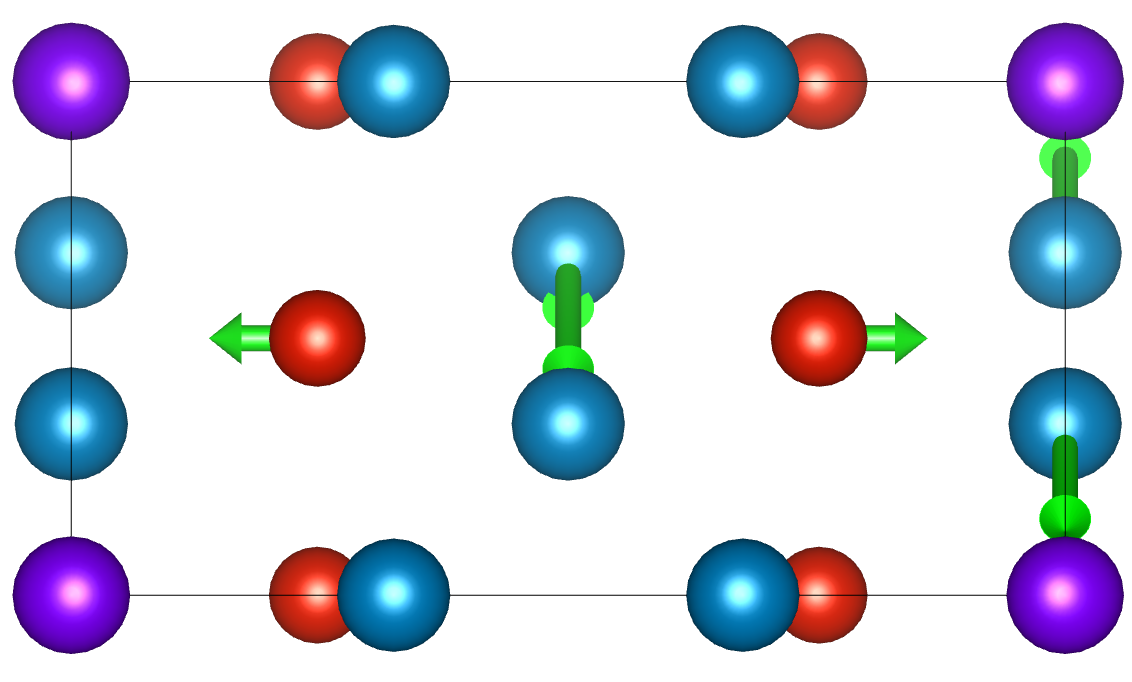} }& \raisebox{17pt} {\includegraphics[width=0.2\linewidth] {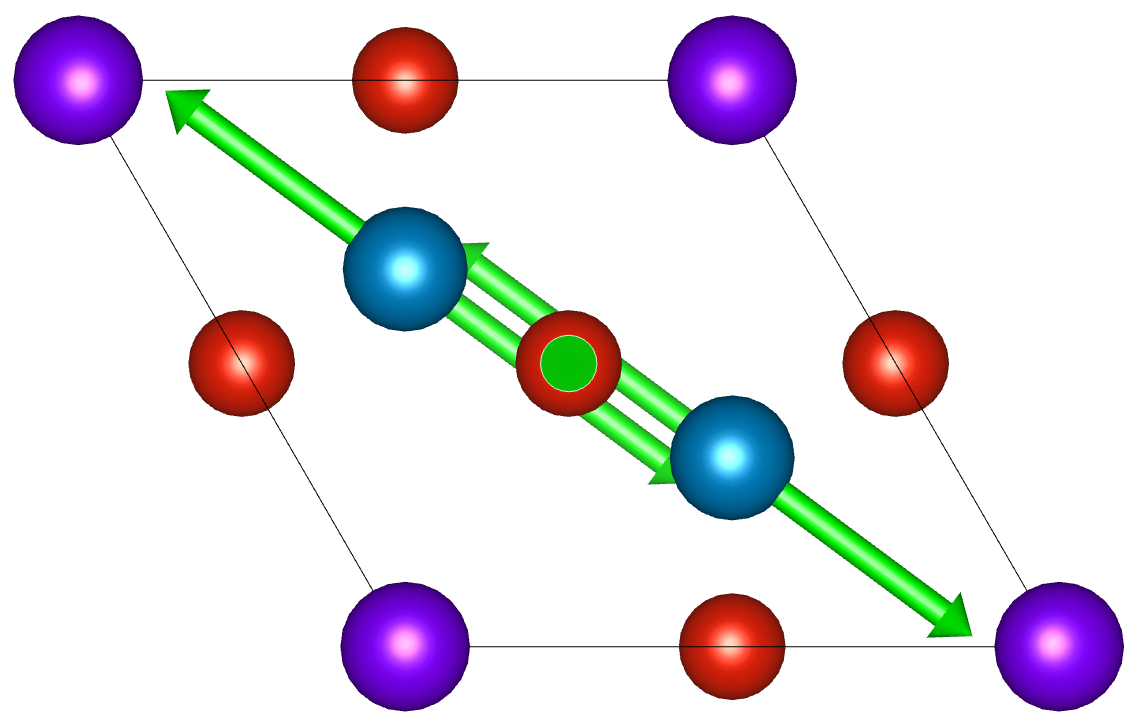}}\\
\hline

\multirow{-12} *{$E_{2g}$} & \multirow{-12} *{140.9}  & \multirow{-12} *{142} &  \includegraphics[width=0.12\linewidth] {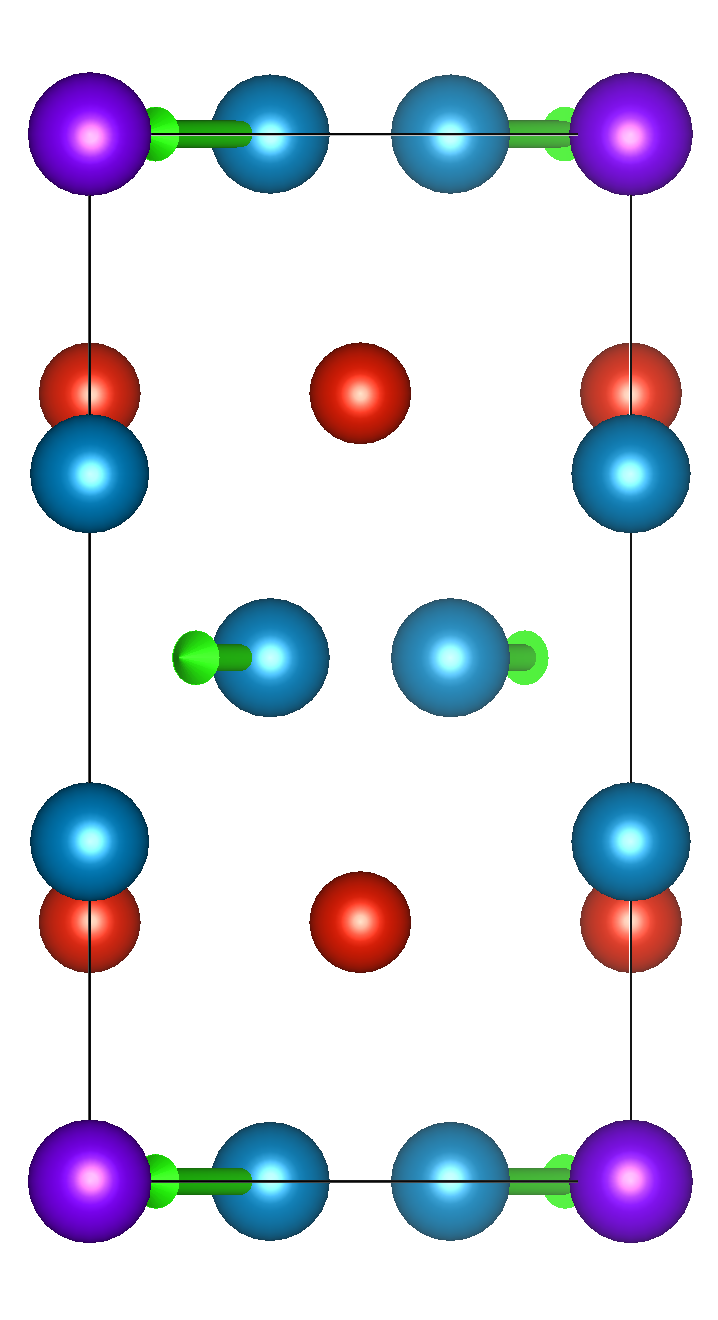} & \raisebox{20pt} {\includegraphics[width=0.2\linewidth] {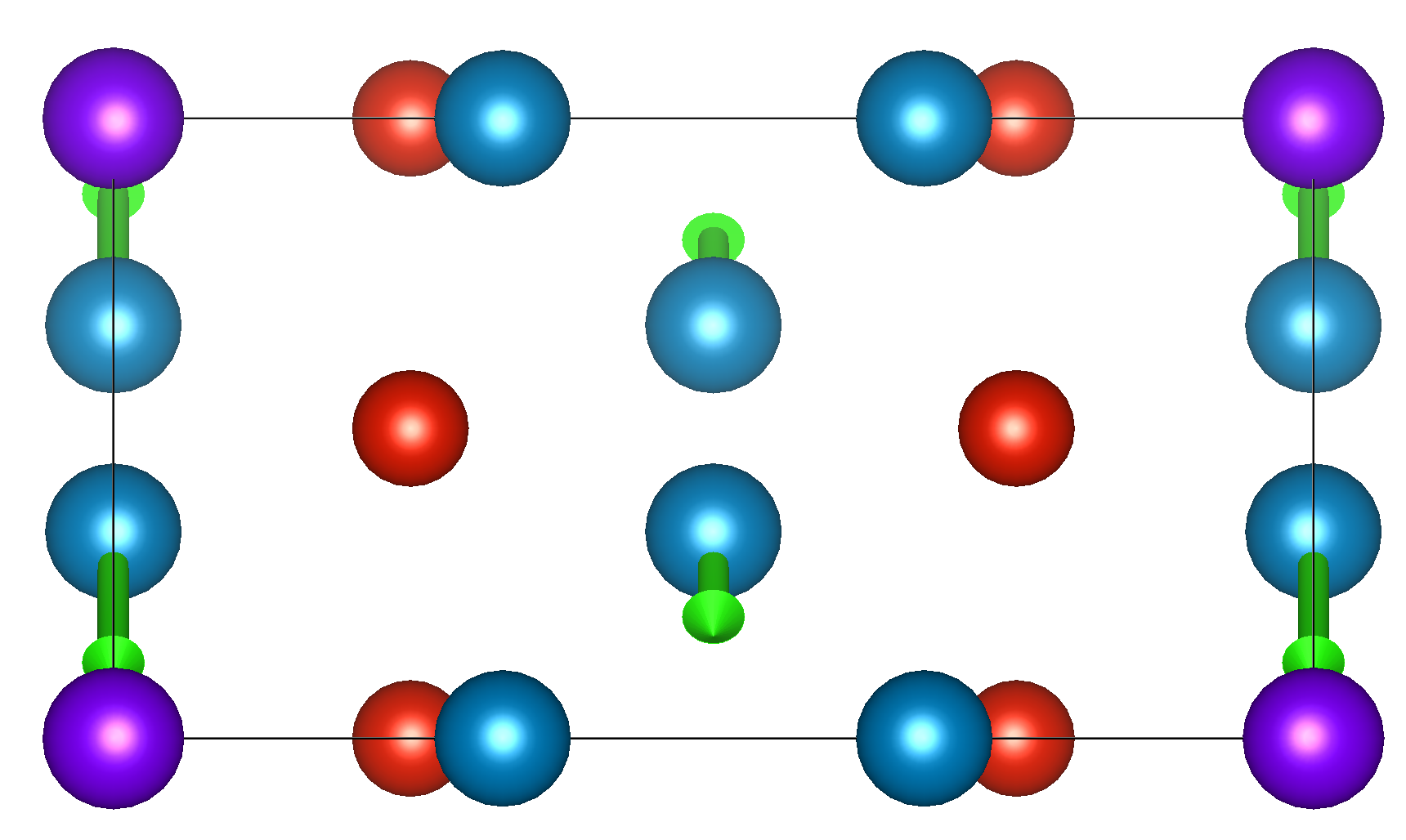}} &\raisebox{20pt} {\includegraphics[width=0.2\linewidth] {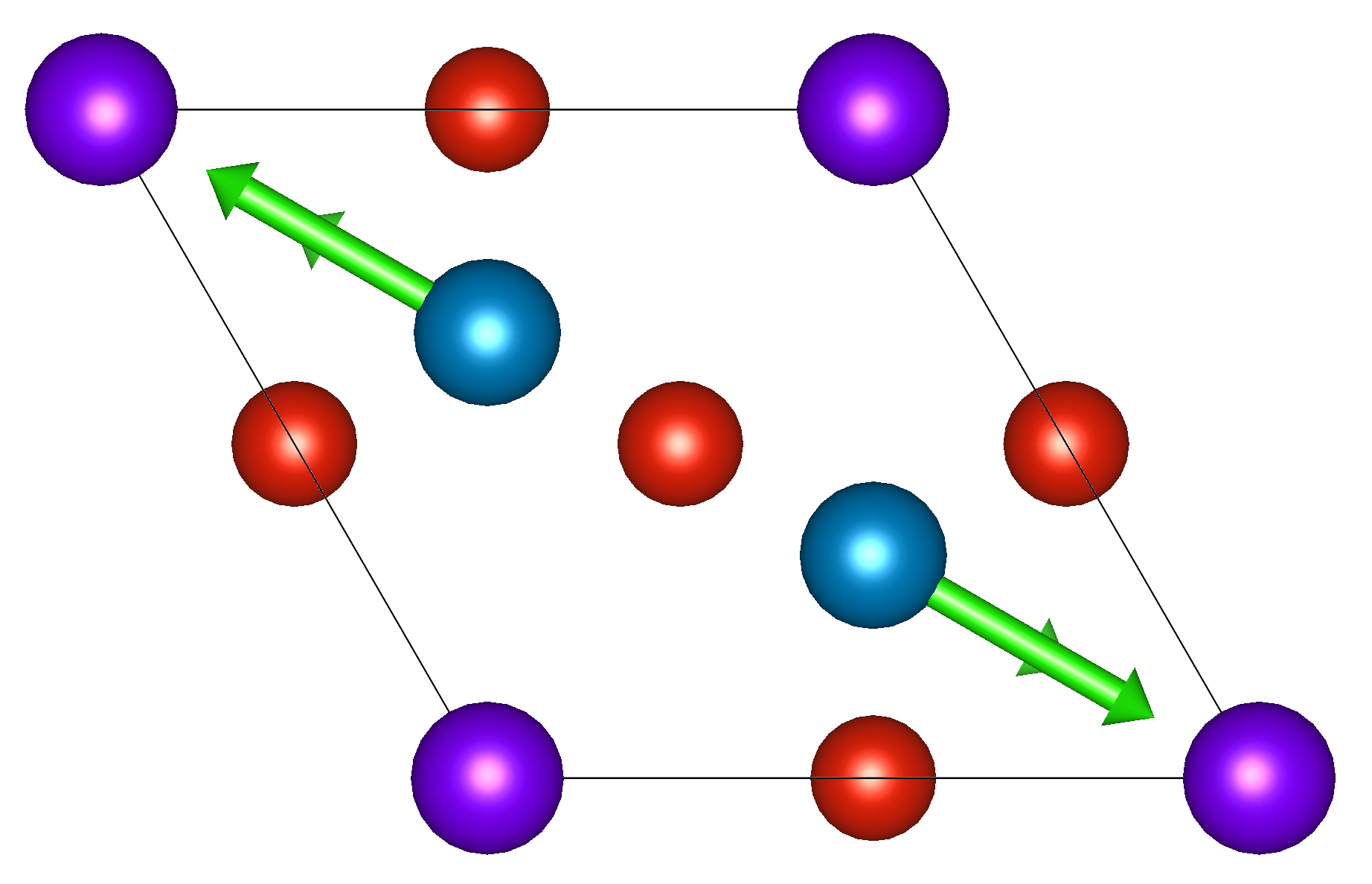}}\\
 \hline
 
  \multirow{-12} *{$E_{1g}$} & \multirow{-12} *{169.5} & \multirow{-12} *{171} & \includegraphics[width=0.2\linewidth] {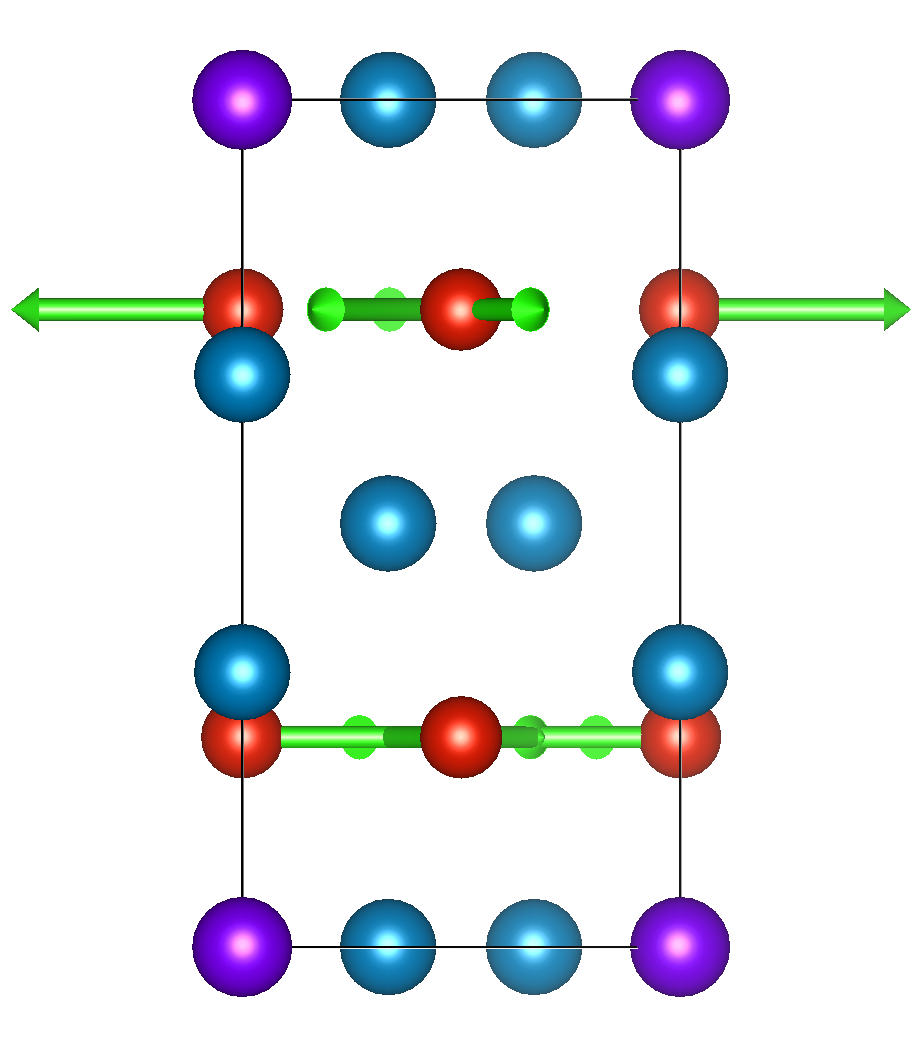} & \raisebox{18pt} {\includegraphics[width=0.2\linewidth] {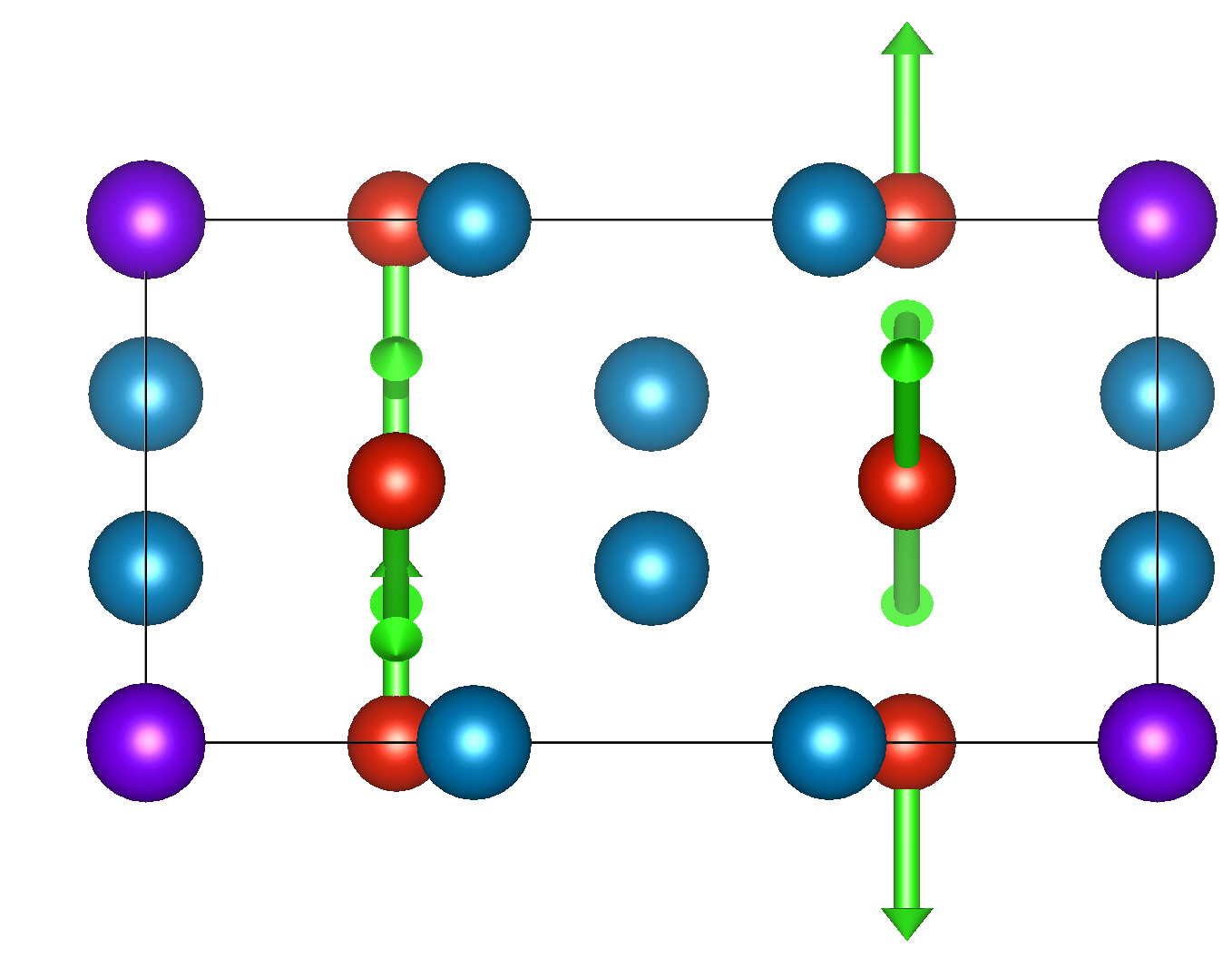} }& \raisebox{18pt} {\includegraphics[width=0.2\linewidth] {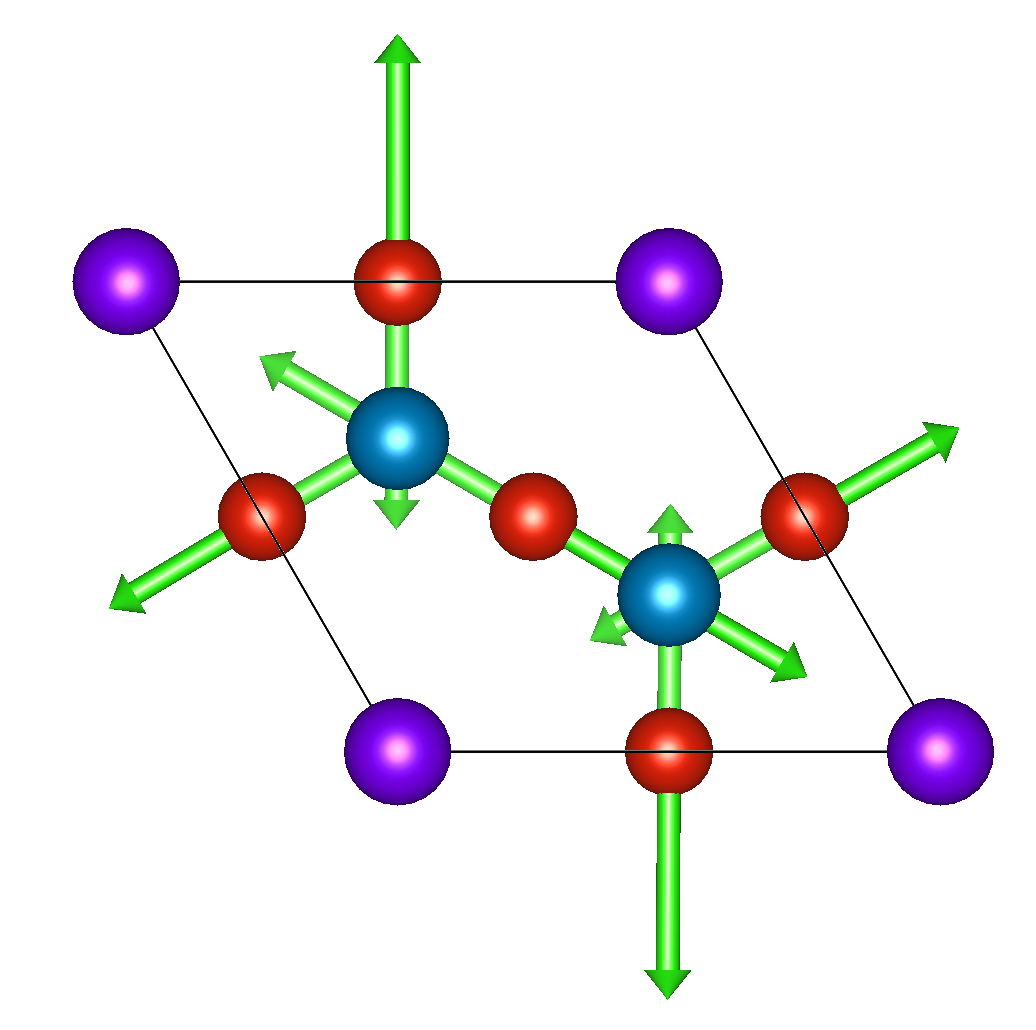}}\\
\hline

  \multirow{-12} *{$E_{1g}$} & \multirow{-12} *{232} & \multirow{-12} *{226} &  \includegraphics[width=0.16\linewidth] {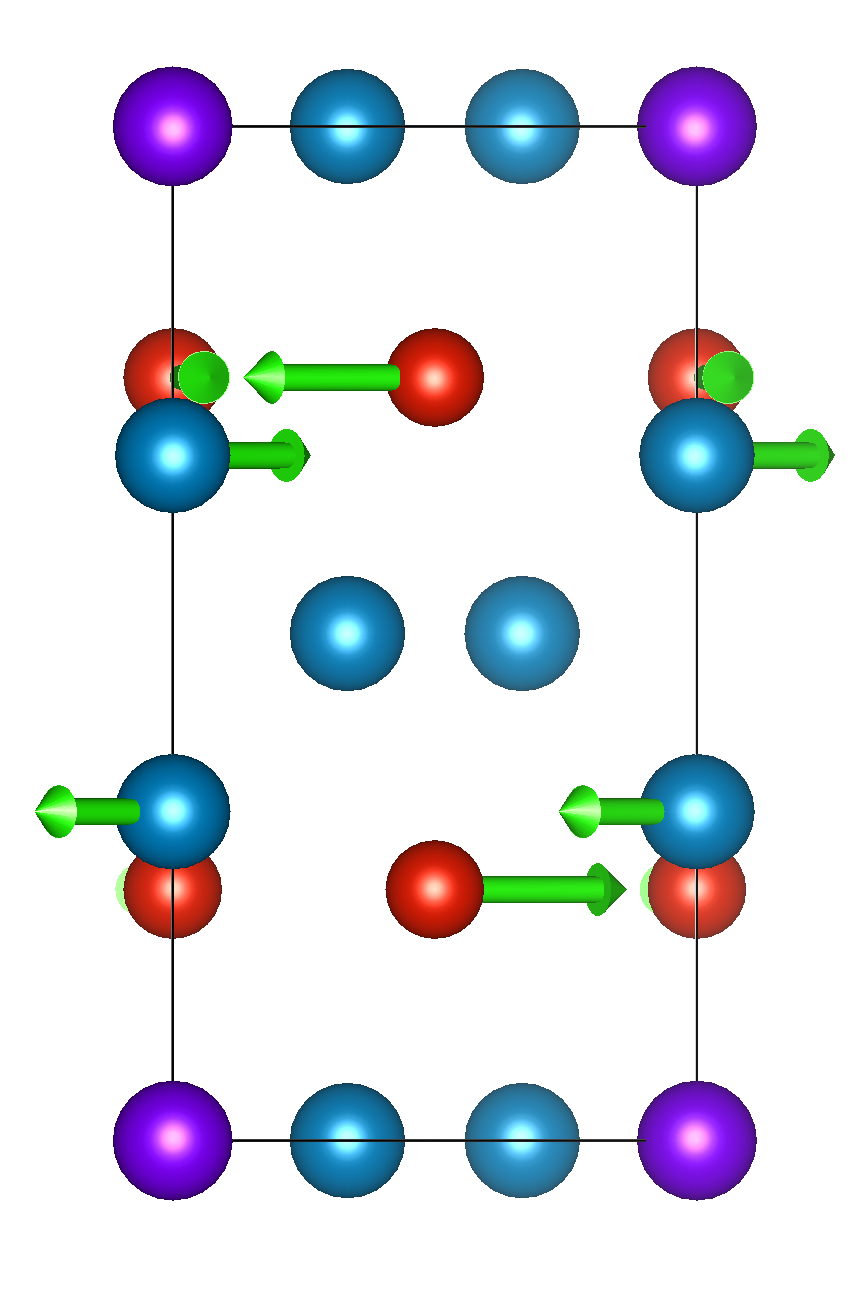} & \raisebox{18pt} {\includegraphics[width=0.2\linewidth] {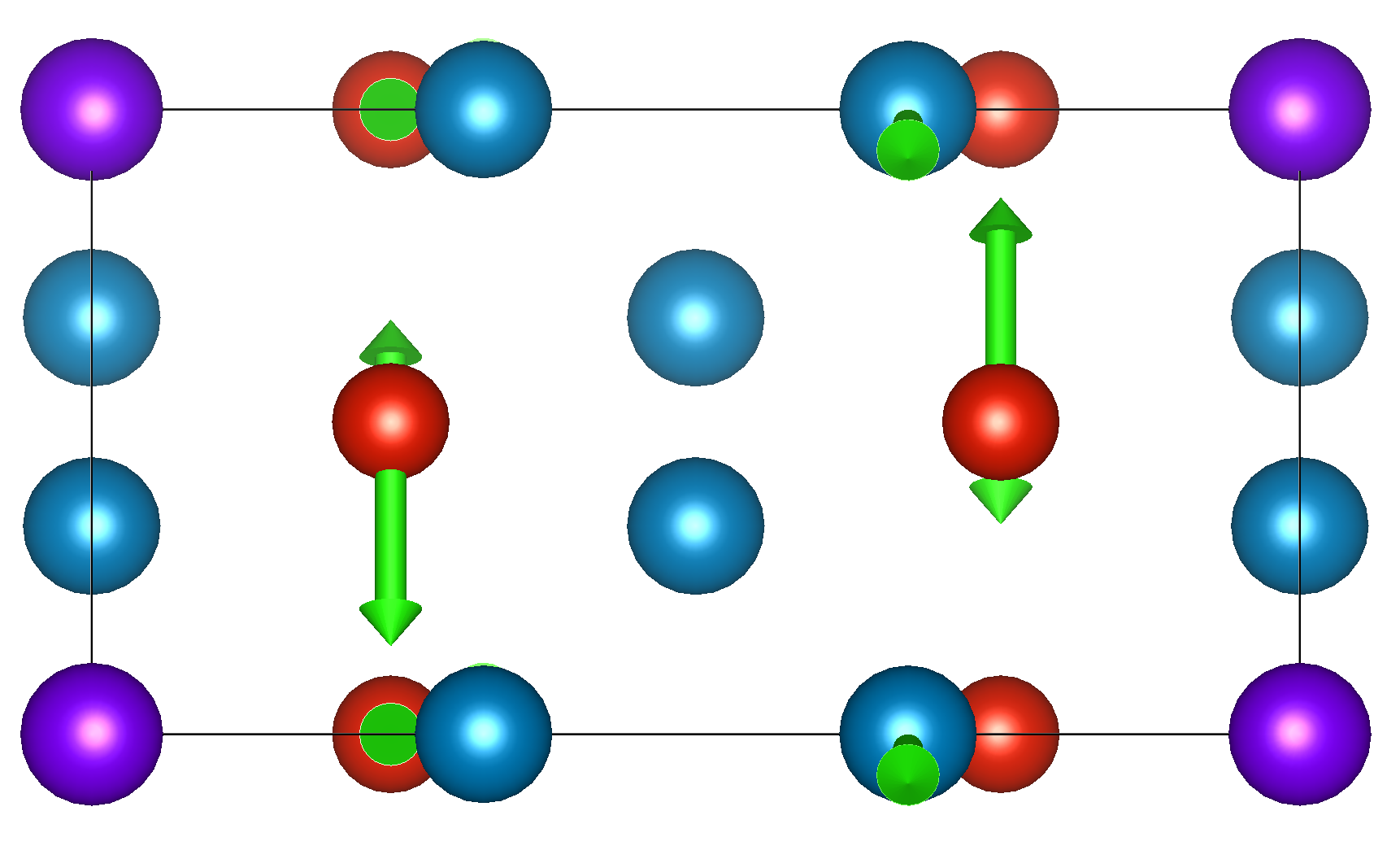}} & \raisebox{18pt} {\includegraphics[width=0.2\linewidth] {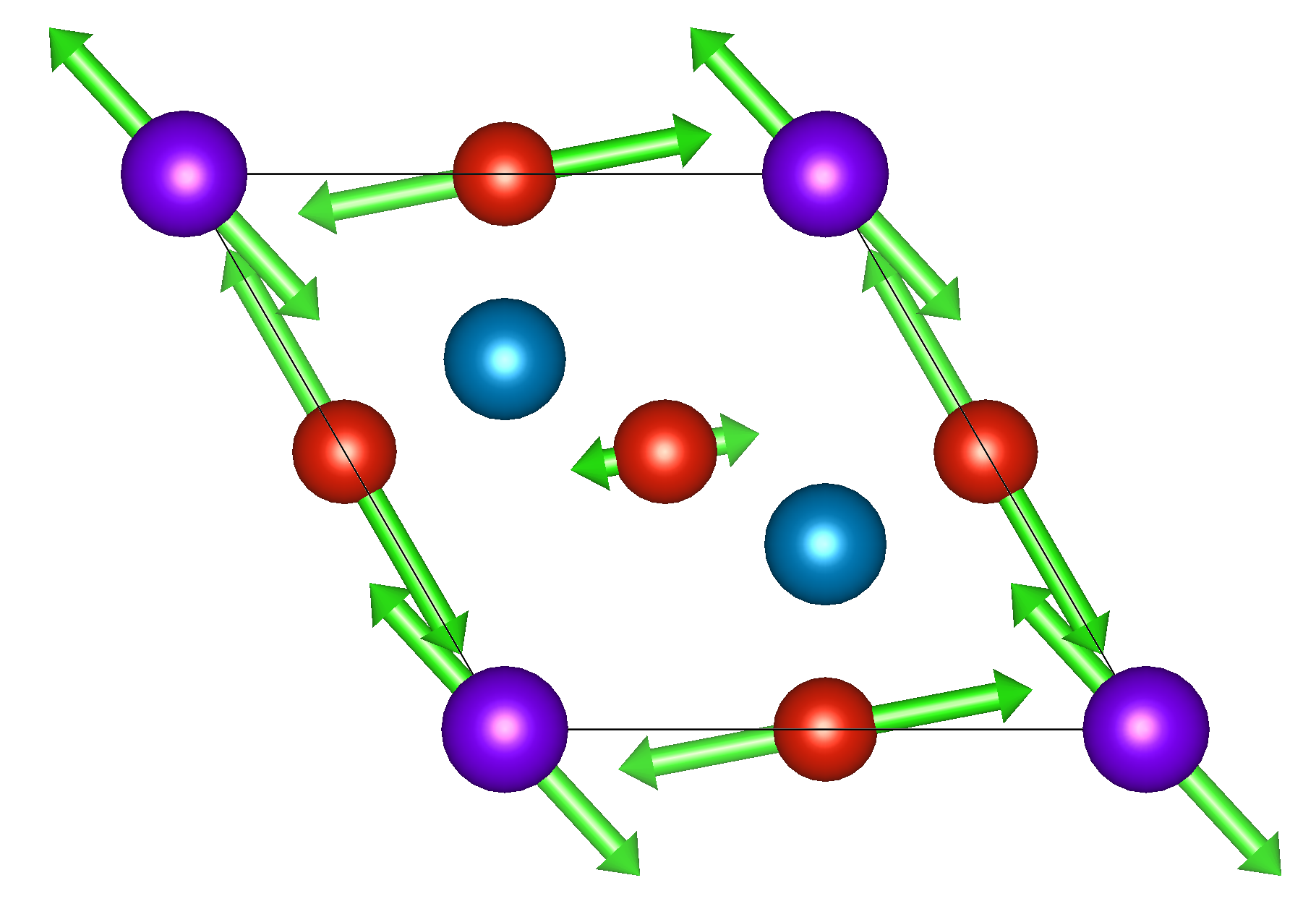}}\\
\hline 

\multirow{-12} *{$A_{1g}$} & \multirow{-12} *{239.9}  & \multirow{-12} *{235} &  \includegraphics[width=0.14\linewidth] {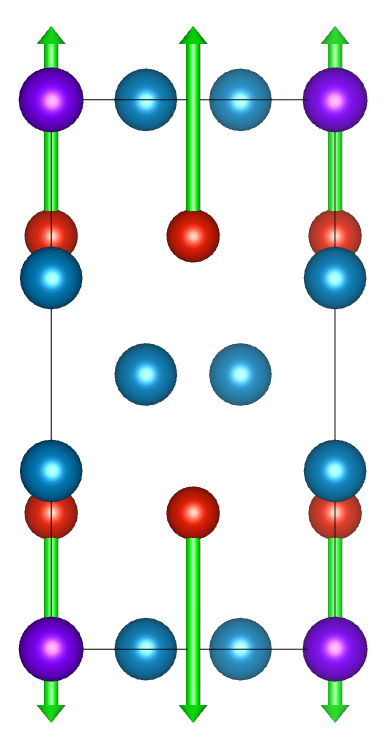} & \raisebox{28pt} {\includegraphics[width=0.2\linewidth] {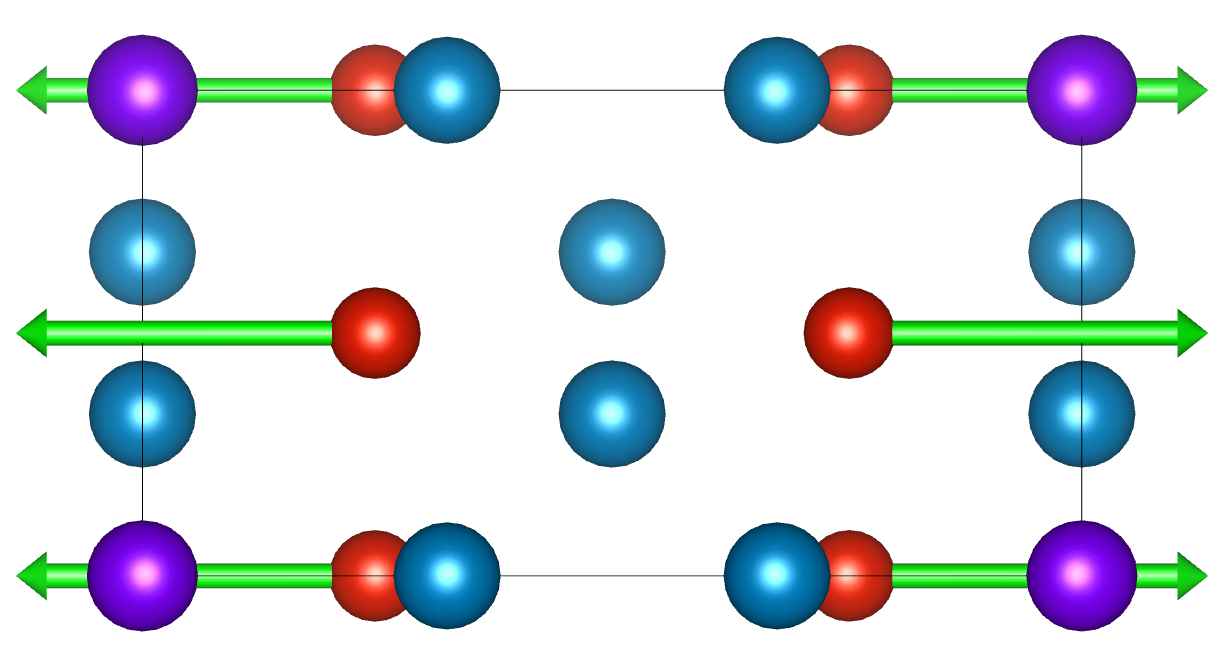} }&\raisebox{28pt} {\includegraphics[width=0.2\linewidth] {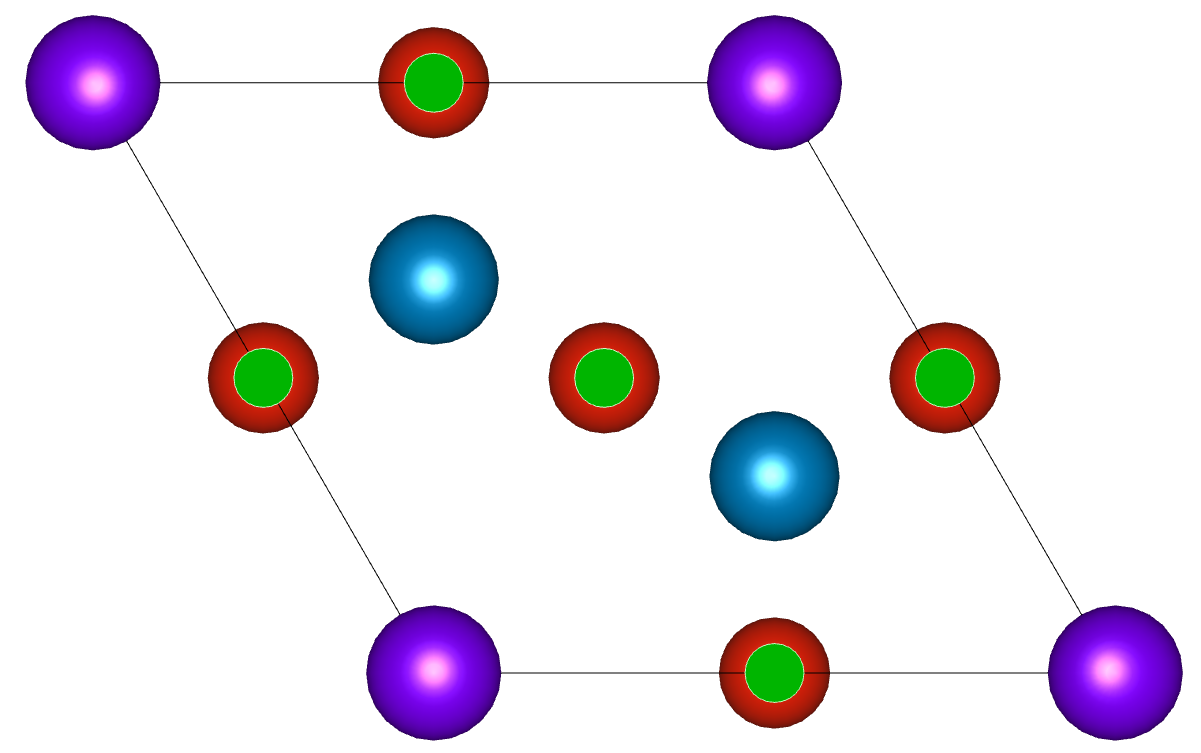}}\\
 \hline
 
\multirow{-12} *{$E_{2g}$} & \multirow{-12} *{301.6}  & \multirow{-12} *{310} &  \includegraphics[width=0.15\linewidth] {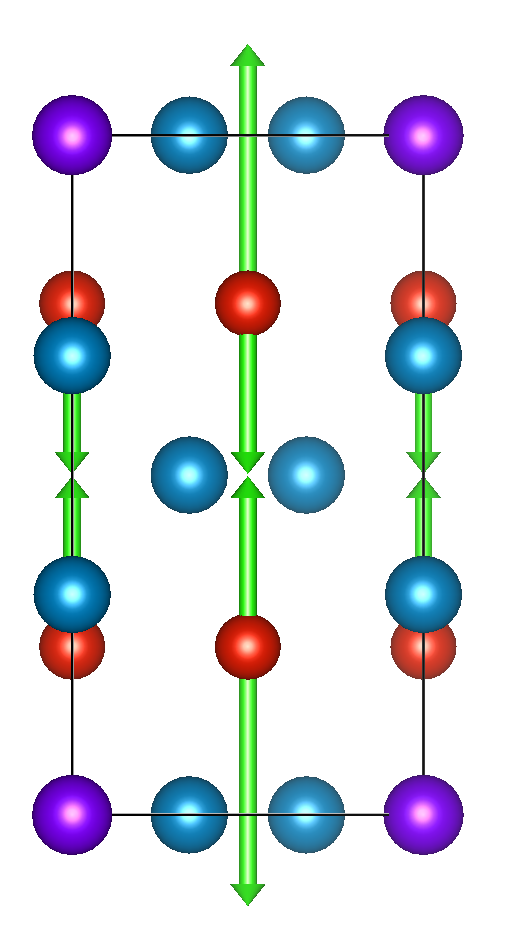} & \raisebox{35pt} {\includegraphics[width=0.2\linewidth] {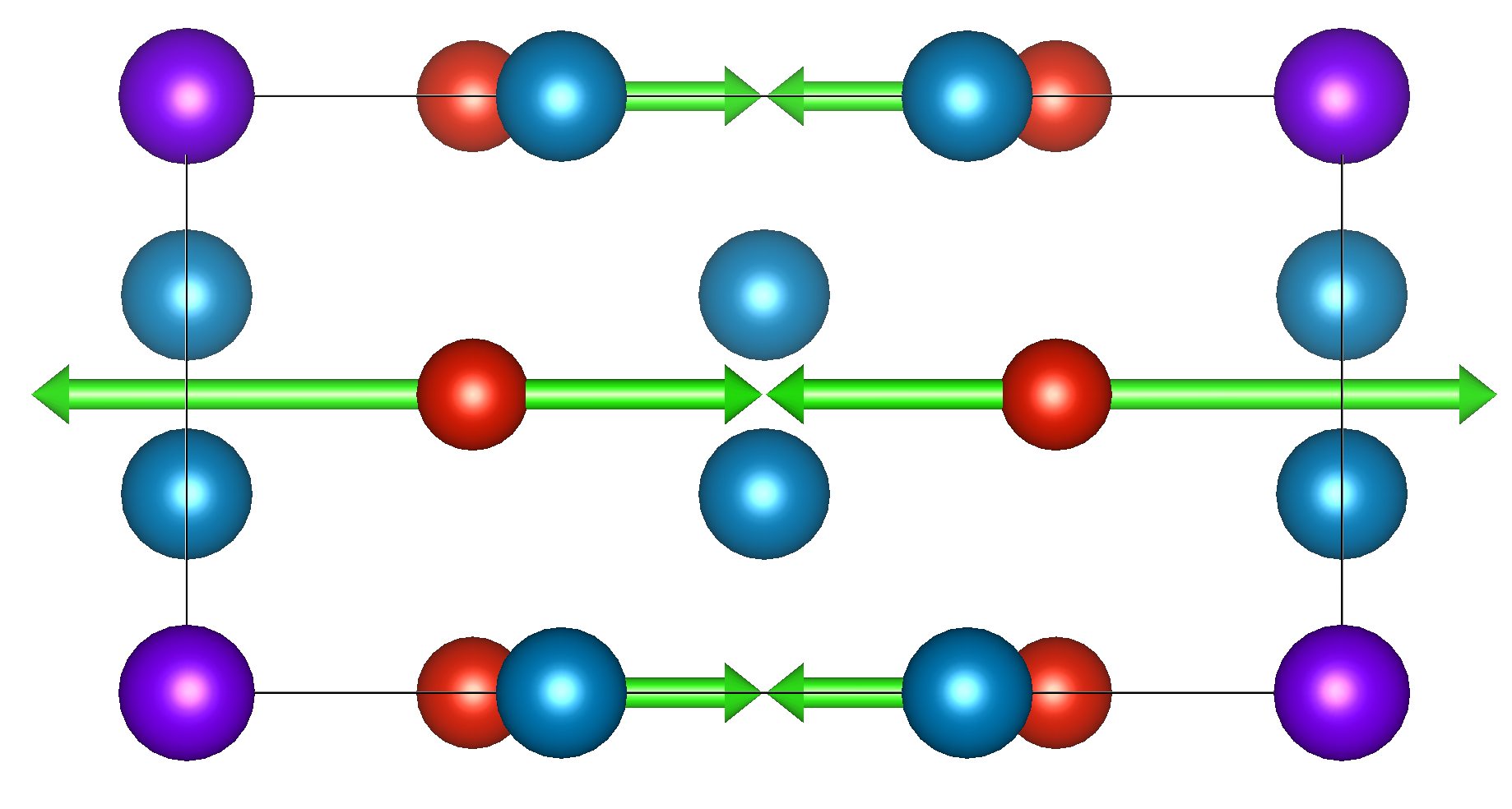}} &\raisebox{35pt} {\includegraphics[width=0.2\linewidth] {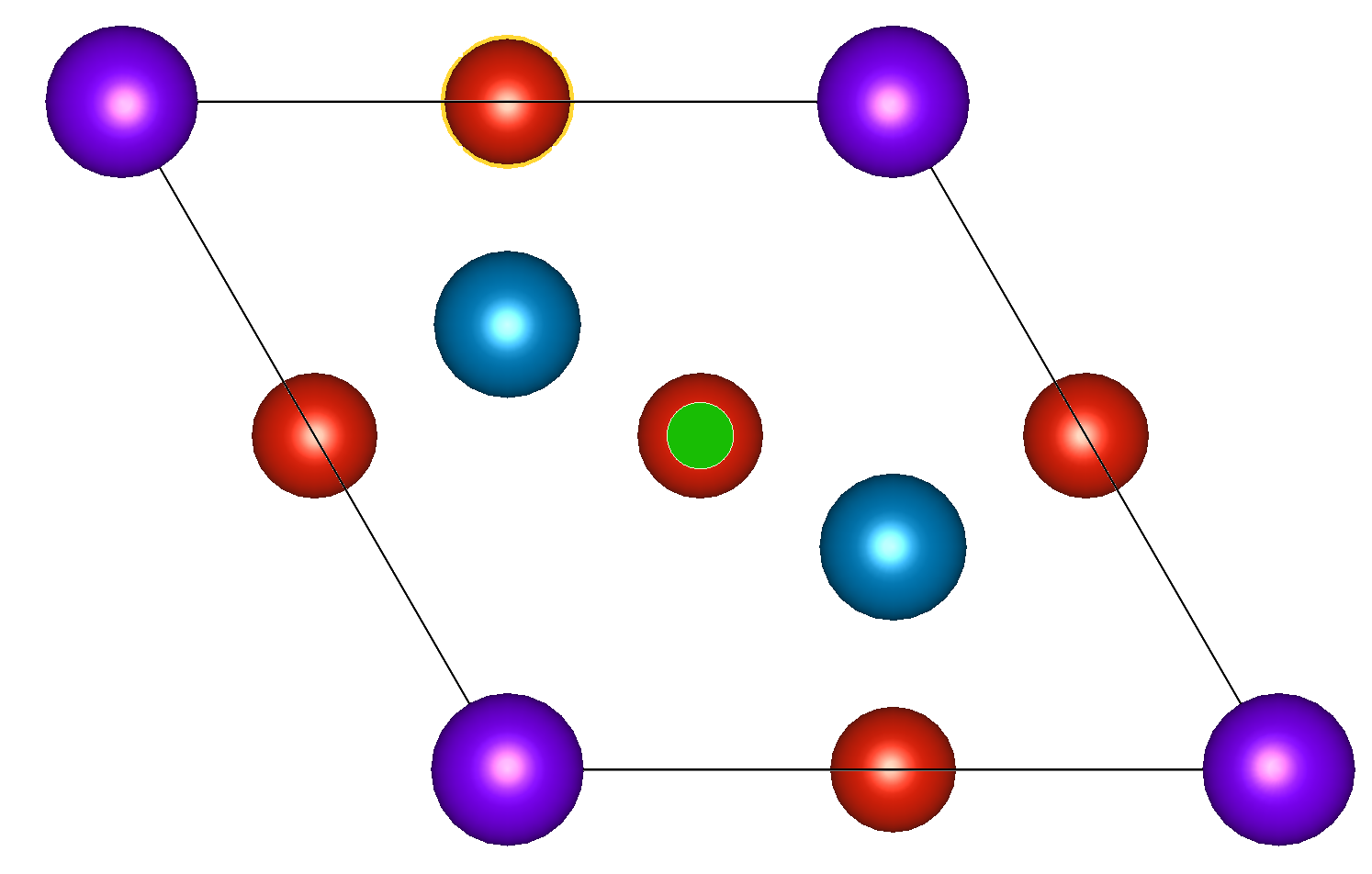}}\\
 \hline

\end{longtable}

\section{SUPPLEMENTARY NOTE 2: Raman scattering response of LuV$_6$Sn$_6$ and YV$_6$Sn$_6$ as a function of temperature}

To verify whether multiplet splitting of the  $A_{1g}$ phonon in ScV$_6$Sn$_6$ is related to the development of the CDW, we performed Raman scattering measurements as a function of temperature on single crystals of $R$V$_6$Sn$_6$($R$ = Lu, Y). This is a good test because  no CDW order had been observed in the Y and Lu analogs \cite{zhang2022electronic,EPokharel2021prb}. The spectra of LuV$_6$Sn$_6$ and YV$_6$Sn$_6$ are shown in Supplementary Figure 2\textbf{a,b}. The low frequency modes are very similar in 
 $R$ = Sc, Lu and Y compounds at both low and high temperatures. There is no splitting near $A_{1g}$ mode at low temperature in either LuV$_6$Sn$_6$ and YV$_6$Sn$_6$ compounds, consistent with our conclusion that splitting of the $A_{1g}$ phonon is a signature of the CDW in ScV$_6$Sn$_6$. Interestingly, a small shoulder is observed near the $A_{1g}$ phonon peak in LuV$_6$Sn$_6$, in line with the prediction of a low intensity $E_{1g}$ symmetry peak. This feature is small and not detected in the other samples.

\begin{figure*}[t!]
\centering
\includegraphics[width=\linewidth]{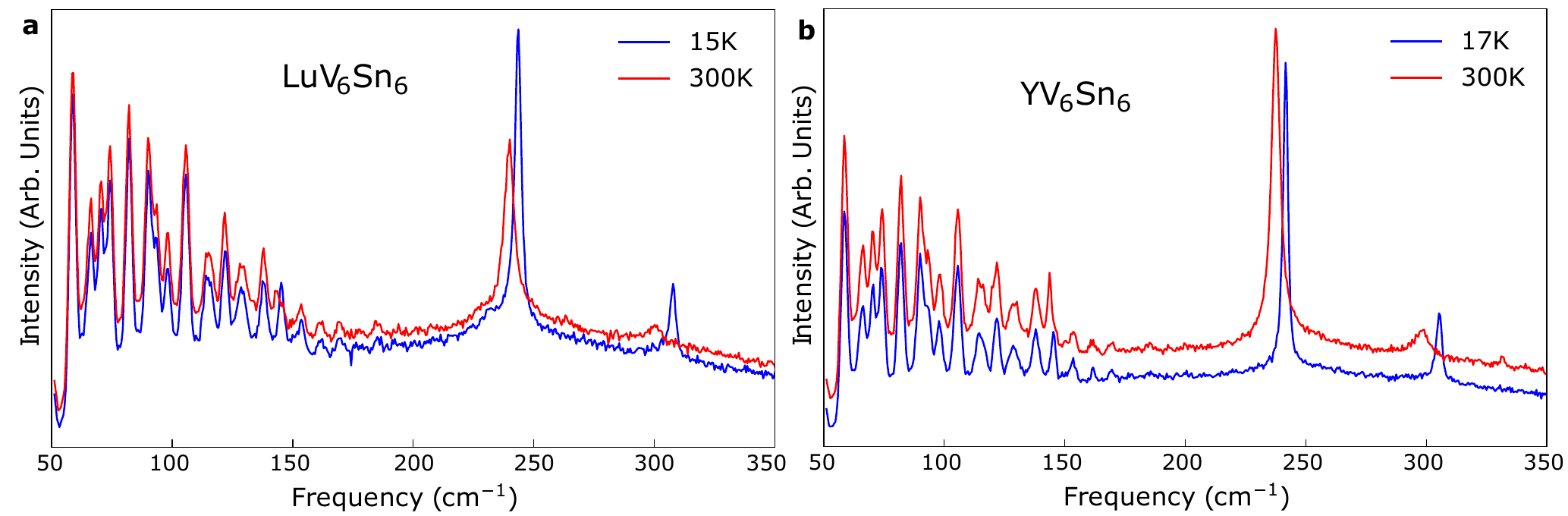}
\caption{\textbf{a} Raman scattering response collected at  15 K (blue) and 300 K (red) on LuV$_6$Sn$_6$. \textbf{b} Raman spectra measured at 17 K (blue) and  300 K (red) on YV$_6$Sn$_6$. Neither shows multiplet splitting of the mode near 240 cm$^{-1}$.}
\label{fig:3}
\end{figure*}

\begin{figure*}[tbh]
\centering
\includegraphics[width=\linewidth]{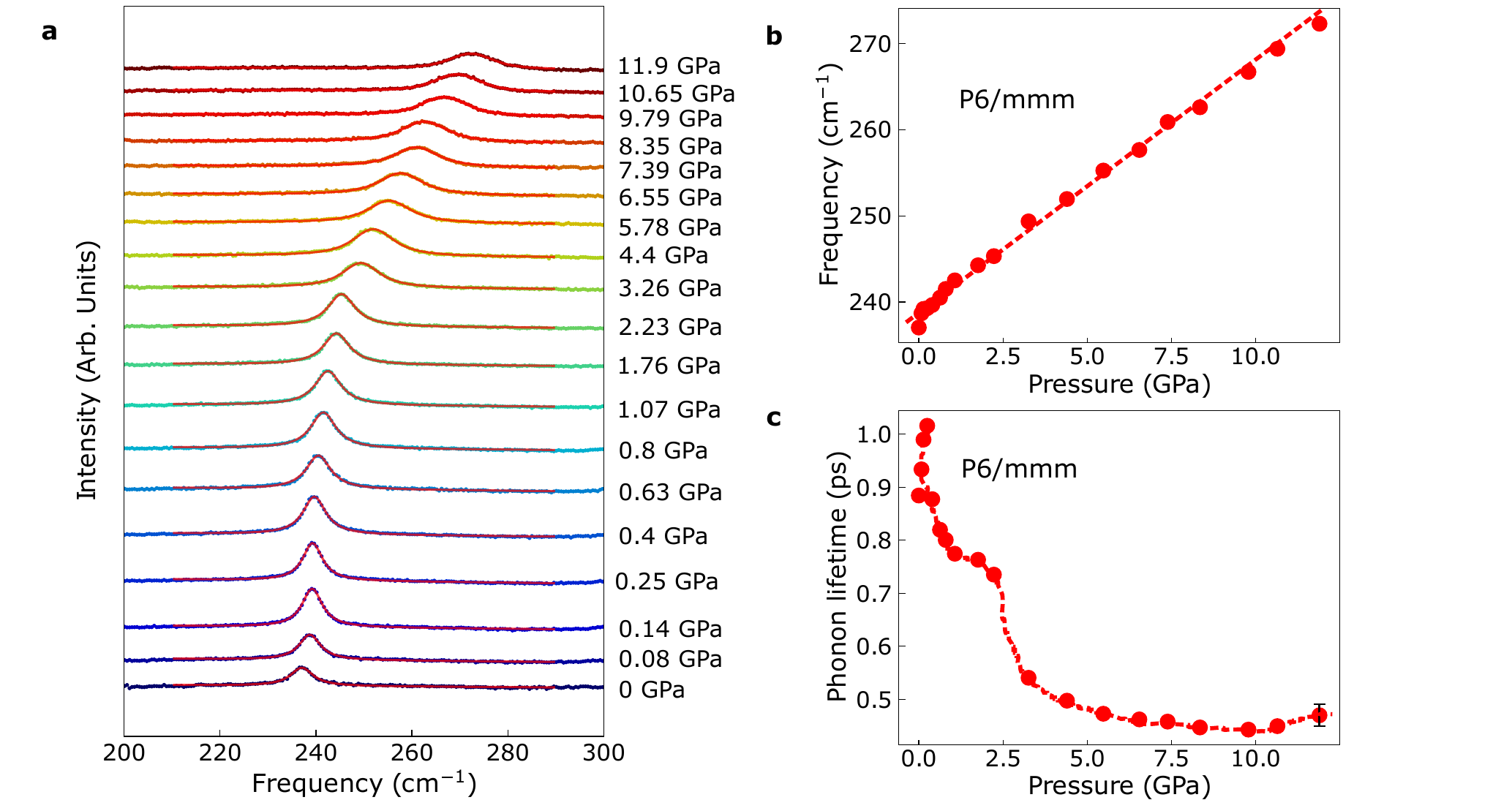}
\caption{\textbf{a}  Raman spectra of LuV$_6$Sn$_6$ from 0 to 11.07 GPa at room temperature. The dots are experimental data, and the solid lines are fits to a single Voigt oscillator with a linear background. \textbf{b} Raman frequency shift as a function of pressure. This mode hardens steadily with increasing pressure. \textbf{c} Pressure dependence of the phonon lifetime. The error bars in panel \textbf{b} are on the order of the symbol size; a characteristic error bar is indicated in panel \textbf{c}.
}
\label{supfig:4}
\end{figure*}

\section{SUPPLEMENTARY NOTE 3: Raman scattering of LuV$_6$Sn$_6$ under pressure}

As discussed in the main text, Sc has the smallest radius compared to Lu and Y. This difference in chemical pressure likely explains why there is a charge density wave transition in ScV$_6$Sn$_6$ but not in the Lu and Y analogs. ScV$_6$Sn$_6$ also hosts a pressure-driven transition that we attribute to quenching of the remnant density wave at room temperature. The ability of pressure to quench short range density wave tendencies is also discussed in the main text. 

To ascertain whether this behavior is present in other $R$V$_6$Sn$_6$ compounds, we performed similar measurements on LuV$_6$Sn$_6$. 
 Supplementary Figure \ref{supfig:4}\textbf{a} displays the $A_{1g}$ phonon peak under pressure ranging from 0 GPa at the bottom to 11.9 GPa at the top. All data were collected at room temperature.  The $A_{1g}$ phonon retains a very good single peak shape from ambient conditions to 11.9 GPa, eliminating the possibility of a pressure-induced transition up to 11.9 GPa. This peak can be modeled with a single oscillator across the entire pressure range. 
The extracted Raman frequency and life time under pressure are shown in Supplementary Figure \ref{supfig:4}\textbf{b,c}. The $A_{1g}$ phonon peak position shifts linearly and systematically under compression. There is thus no evidence for a pressure-driven transition in LuV$_6$Sn$_6$. The phonon lifetime is different. As discussed in the main text, this quantity is derived from the linewidth, and examination of the data in Supplementary Figure \ref{supfig:4}\textbf{a} reveals an obvious increase in linewidth near 3 GPa. So while we can see that the overall phonon life time decreases from approximately 1 to 0.5 ps under compression, there is a significant drop near 3 GPa. Even so, it is clearly much more difficult to induce a phase transition under a similar pressure range as we observed in ScV$_6$Sn$_6$ - probably due to the larger radius in Lu compared to Sc. These differences are highlighted in Supplementary Figure \ref{supfig:5}.  In the LuV$_6$Sn$_6$ case, a higher pressure may be needed to achieve a new ground state. On the other hand, the absence of a pressure-driven transition in LuV$_6$Sn$_6$ may be related to the lack of short range CDW in this material. If there is no short range order, there is no correlation to be destroyed in this pressure range.

\begin{figure*}[tbh]
\centering
\includegraphics[width=6.0in]{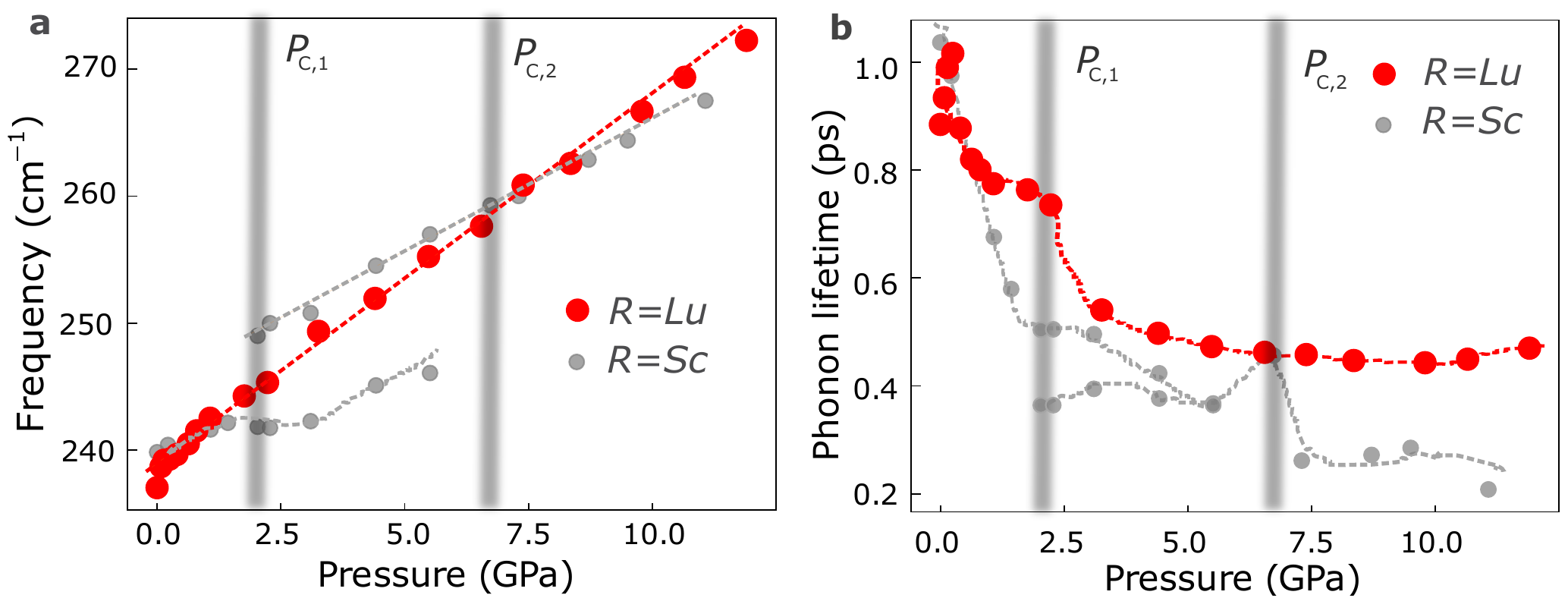}
\caption{\textbf{a} Comparison of the Raman frequency shift vs. pressure for both LuV$_6$Sn$_6$ (red dots) and ScV$_6$Sn$_6$ (gray dots). \textbf{b} Comparison of the phonon lifetime of LuV$_6$Sn$_6$ (red dots) and ScV$_6$Sn$_6$ (gray dots) vs. pressure. The dots are experimental data; the solid lines guide the eye.
}
\label{supfig:5}
\end{figure*}

\section{SUPPLEMENTARY NOTE 4: Predicted phonon frequencies from density functional theory}

The following phonon frequencies were predicted in DFT using the techniques outlined in the Methods section of the main body of the text. Both high and low temperature phases are tabulated. Irrep labels $P_i$ refer to the three irreps at the $P=(\frac{1}{3}, \frac{1}{3}, \frac{1}{3})$ point in reciprocal space; our definition of these irreps in terms of the characters of the generators of the little group of $P$ in $P6/mmm$ is included in Table \ref{tab:irrepLabelP} (our notation matches that found on the Bilbao Crystallographic Server \cite{aroyo2006}.

\begin{table}[h!]
\begin{tabular}{|c|c|c|c|c|c|c|c|c|c|c|}
\hline
\multicolumn{1}{|c}{\text{Irrep}}  & \multicolumn{1}{|c}{Activity} & \multicolumn{9}{|c|}{Frequency (cm$^{-1}$)}\\[0.124cm]
\hline
$\Gamma_1^+$ / A$_{1g}$ & Raman & 112  & 235 & \multicolumn{7}{c}{}\\
\cline{1-5}
$\Gamma_3^+$ / B$_{2g}$&  & 78 & 177 & 235 &\multicolumn{6}{c}{} \\[0.024cm]
\cline{1-5}
$\Gamma_4^+$ / B$_{1g}$ &  & 272  & \multicolumn{8}{c}{}\\
\cline{1-5}
$\Gamma_5^+$ / E$_{2g}$ & Raman & 124 & 142 & 310 & \multicolumn{6}{c}{}\\
\cline{1-5}
$\Gamma_6^+$ / E$_{1g}$ & Raman & 83 & 171 & 226 & \multicolumn{6}{c}{}\\
\cline{1-7}
$\Gamma_2^-$ / A$_{2u}$ & IR & 0 & 91 & 173 & 217 & 263 & \multicolumn{4}{c}{}\\
\cline{1-7}
$\Gamma_3^-$ / B$_{2u}$  & & 271 & \multicolumn{8}{c}{}\\
\cline{1-3}
$\Gamma_4^-$ / B$_{1u}$  & & 142 & \multicolumn{8}{c}{}\\
\cline{1-3}
$\Gamma_5^-$ / E$_{2u}$  & & 255 & \multicolumn{8}{c}{}\\
\cline{1-8}
$\Gamma_6^-$ / E$_{1u}$  & IR & 0 & 82 & 102 & 176 & 181 & 232 & \multicolumn{3}{c}{} \\
\cline{1-11} 
$P_1$  &  & 12$i$ & 105 & 129 & 147 & 220 & 229 & 234 & 272 & 286 \\
\cline{1-11}
$P_2$  &  & 77 & 98 & 186  & 191 & \multicolumn{5}{c}{} \\
\cline{1-11}
\multirow{2}{*}{$P_3$}  &  & 100 & 102 & 128 & 133 & 137 & 146 & 180 & 216 & 221  \\
\cline{3-11}
 & & 227 & 235 & 266 & 269 & \multicolumn{5}{c}{}\\
 \cline{1-6}
\end{tabular}
\centering
\caption{Predicted phonon frequencies at wavevectors $\Gamma$ and $P=(\frac{1}{3}, \frac{1}{3}, \frac{1}{3})$ for the $P6/mmm$ structure, as well as their Raman and IR activity, grouped by irreducible representation. }
\label{tab:191phon}
\end{table}

\begin{table}[tbh]
      \caption{ Characters of the irreps of the little group of the $P6/mmm$ $P$ point. We only list the characters for the three (non-translational) generators of the little group, as given by the Bilbao Crystallographic Server \cite{aroyo2006}.}
    \label{tab:irrepLabelP}
    \centering
    \begin{tabular}{|c|c|c|c|}
    \hline
        Irrep & $\mathbf{1}$ & $3^+_{001} $  & $m_{1\bar{1}0}$ \\ \hline
       $P_1$ & $+1$ & $+1$ & $+1$  \\
       $P_2$ & $+1$ & $+1$ & $-1$   \\
        $P_3$ & $+2$ & $-1$ &  $0$  \\
       \hline
       \end{tabular}
\end{table}

\begin{table}[tbh]
      \caption{ Characters of the irreps of the $P6/mmm$ $\Gamma$ point. We only list the characters for the non-translational generators of the group as given by the Bilbao Crystallographic Server \cite{aroyo2006}.}
         \label{tab:irrepLabelGM}
    \centering
    \begin{tabular}{|c|c|c|c|c|c|}
    \hline
        Irrep & $\mathbf{1}$ & $3^+_{001} $  & $2_{001}$ & $2_{110}$ & $\mathbf{-1}$\\ \hline
        $\Gamma_1^{+}$ & $+1$ & $+1$ & $+1$ &  $+1$  &  $+1$\\
        $\Gamma_2^{-}$ & $+1$ & $+1$ & $+1$ & $-1$ &  $-1$  \\
        $\Gamma_3^{\pm}$  & +1$ $& $+1$ & $-1$ & $+1$ & $\pm1$    \\
        $\Gamma_4^{\pm}$  &  $+1$ & $+1$ & $-1$ & $-1$ & $\pm1$    \\
        $\Gamma_5^{\pm}$  & $+2$& $-1$ & $+2$ & $0$ & $\pm 2 $   \\
        $\Gamma_6^{\pm}$  & $+2$& $-1$ & $-2 $& $0$ & $\pm 2 $  \\
        \hline
       \end{tabular}
\end{table}

\begin{table}[h!]
\begin{tabular}{|l|c|c|c|c|c|c|c|c|c|c|c|c|c|c|c|c|c|c|c|c|c|}
\hline
\multicolumn{1}{|c}{\text{Irrep}}  & \multicolumn{1}{|c}{Activity} & \multicolumn{20}{|c|}{Frequency (cm$^{-1}$)}\\[0.124cm]
\hline
$\Gamma_1^+$ / A$_{1g}$ & Raman & 44 & 107 & 118 & 133 & 144 & 232 & 237 & 240 & \multicolumn{12}{c}{}\\
\cline{1-10}
$\Gamma_2^+$ / A$_{2g}$& & 80 & 83 & 99 & 180 & 191 & 196 &240 &  \multicolumn{13}{c}{} \\
\cline{1-21}
$\Gamma_3^+$ / E$_{g}$& Raman  & 84 & 102 & 102 & 126 & 131 & 133 & 143 & 145 & 146 & 175 & 181 & 220 & 228 & 232 & 237 & 239 & 272 & 276 & 314 & \multicolumn{1}{c}{} \\
\cline{1-21}
$\Gamma_1^-$ / A$_{1u}$&  & 76 & 99 & 150  & 190 & 195 & \multicolumn{14}{c}{} \\
\cline{1-17}
$\Gamma_2^-$ / A$_{2u}$& IR & 0 & 66 & 93 & 114 & 136 & 146 & 176 & 209 & 230 & 234 & 243 & 269 & 274 & 275 & 294 & \multicolumn{5}{c}{} \\
\cline{1-22}
$\Gamma_3^-$ / E$_{u}$& IR & 0 & 83 & 103 & 104  & 105 & 131 & 134 & 143 & 145 & 178 & 183 & 184 & 222 & 231 & 233 & 237 & 242 & 262 & 273  & 274 \\
\cline{1-22}

\end{tabular}
\centering
\caption{Predicted phonon frequencies at wavevector $\Gamma$  for the $R\bar{3}m$ structure, as well as their Raman and IR activity, grouped by irreducible representation. }
\label{tab:191phon}
\end{table}

 



\nocite{*}
\clearpage
\bibliography{ref_sup}